\documentclass[
 aip,
 jcp,
 amsmath,amssymb,
 reprint,
]{revtex4-1}

\usepackage{graphicx}
\usepackage{dcolumn}
\usepackage{bm}
\usepackage[utf8]{inputenc}
\usepackage[T1]{fontenc}
\usepackage{mathptmx}
\usepackage{etoolbox}
\usepackage{mathtools}
\usepackage{mathrsfs}
\usepackage{amsthm}
\usepackage{physics}
\usepackage{xcolor}
\usepackage{enumitem}
\usepackage{microtype}
\usepackage{hyperref}
\hypersetup{hidelinks}

\makeatletter
\def\@email#1#2{%
 \endgroup
 \patchcmd{\titleblock@produce}
  {\frontmatter@RRAPformat}
  {\frontmatter@RRAPformat{\produce@RRAP{*#1\href{mailto:#2}{#2}}}\frontmatter@RRAPformat}
  {}{}
}%
\makeatother

\begin{document}

\title[]{A unified variational framework for the inverse Kohn--Sham problem}

\author{Nan Sheng}
\affiliation{
Institute for Computational and Mathematical Engineering (ICME), Stanford University, Stanford, CA 94305, USA.
}%
\email{nansheng@stanford.edu}

\date{\today}

\begin{abstract}
The inverse Kohn--Sham (KS) problem seeks a local effective potential whose noninteracting ground state reproduces a prescribed electron density. Although many inversion formulations and schemes have been developed, they are often formulated in disparate languages, including reduced variational optimization, penalty regularization, response-based iteration, and PDE-constrained optimization. In this work, we develop a unified framework for inverse KS theory in two steps. First, we identify the fixed-density noninteracting constrained search and its density--potential duality as the natural variational anchor of the inverse KS problem. In this setting, the KS potential appears as the variational dual object associated with density reproduction, reducing to the familiar multiplier picture in regular regimes. Second, building on this anchor, we classify major inversion formulations according to how the KS state equations and density-reproduction condition are treated within the optimization architecture, with orbital orthonormality retained as an additional structural constraint. Within this framework, the Wu--Yang formulation appears as a potential-space reduced multiplier formulation, the Zhao--Morrison--Parr construction as a quadratic-penalty relaxation, and PDE-constrained approaches as explicit state-constraint formulations at the orbital level. Rather than comparing inversion formulations primarily at the level of implemented algorithms, the present work develops an optimization-theoretic formulation map. This viewpoint identifies where additive-constant ambiguity, asymptotic normalization, nonsmooth variational structure, metric choice, and weak-gap instability enter different inversion architectures, and it makes explicit how major inversion approaches are connected and where algorithmic design choices arise.
\end{abstract}

\maketitle

\section{Introduction}

Density functional theory (DFT) is one of the central frameworks for the electronic structure of interacting many-body systems in chemistry and condensed-matter physics. Its formal basis is provided by the Hohenberg--Kohn theorems \cite{HohenbergKohn1964}, which establish, under appropriate assumptions, a one-to-one correspondence between the ground-state density and the external scalar potential, up to an additive constant. Within the Kohn--Sham (KS) construction \cite{KohnSham1965}, this correspondence motivates the introduction of an auxiliary noninteracting system whose ground-state density matches that of the interacting problem, while exchange and correlation are absorbed into an effective one-body potential.

A persistent question in exact and practical DFT is the \emph{inverse Kohn--Sham problem}: given a prescribed target density $\rho_{\mathrm{tar}}(\mathbf r)$, determine an effective local potential $v_\mathrm{s}(\mathbf r)$ whose noninteracting ground state reproduces that density \cite{ShiWasserman2021,JensenWasserman2018}. In practice, target densities may be obtained from highly accurate electronic-structure calculations, and inversion then serves both as a conceptual probe of the density--potential correspondence and as a practical route to reference effective and exchange-correlation potentials \cite{ShiWasserman2021,KanungoZimmermanGavini2019}. It also provides a useful setting for diagnosing the limitations of approximate functionals and for generating benchmark data for data-driven electronic-structure modeling \cite{KanungoZimmermanGavini2019,ShiChavezWasserman2022,KanungoZimmermanGavini2023}.

Over the past three decades, several influential inversion approaches have been developed, including the Zhao--Morrison--Parr (ZMP) method \cite{ZhaoMorrisonParr1994}, the Wu--Yang variational formulation \cite{WuYang2003}, and PDE-constrained optimization formulations in which the KS equations are retained as explicit state constraints \cite{JensenWasserman2018}. A broader class of inversion and stabilization strategies has since been developed, including variational reconstructions, basis-set-stabilized formulations, and response-based schemes \cite{KadantsevStott2004,GaidukRyabinkinStaroverov2013,ZhangCarter2018,ErhardTrushinGorling2022,Gould2023}. Closely related wavefunction- and reduced-density-matrix-based reconstructions of KS potentials have also been introduced \cite{RyabinkinStaroverov2012,RyabinkinKohutStaroverov2015,CuevasSaavedraAyersStaroverov2015}. These methods are often introduced in rather different mathematical languages, including reduced optimization over potentials, quadratic-penalty formulations, response-based iteration, orbital-level reconstruction, and explicit state-equation approaches \cite{ShiWasserman2021,JensenWasserman2018}. A main goal of the present work is to place these formulations within a unified variational framework. The emphasis is structural: variational and optimization-theoretic ideas provide a common language for relating the principal inversion formulations.

The perspective adopted here has two connected levels. At the first level, we argue that the inverse KS problem should be understood not merely as the inversion of a nonlinear density-to-potential map, but as the recovery of a supporting potential in the fixed-density noninteracting constrained-search problem already embedded in exact DFT. More precisely, inverse KS theory is anchored here in the density-constrained inner noninteracting variational problem that appears within the Levy--Lieb and Lieb formulations of exact DFT \cite{Levy1979,Lieb1983}. In this setting, the KS potential appears naturally as the multiplier associated with density reproduction, so that inverse KS theory is placed directly within the variational architecture of exact DFT rather than outside it. This
interpretation is particularly transparent when exact DFT is organized into
parallel interacting and noninteracting ensemble variational hierarchies:
inverse KS theory then appears as the problem of recovering the supporting
potential of the noninteracting kinetic energy functional at a prescribed density.\cite{sheng2026exact}

At the second level, building on this variational anchor, we develop a broader structural classification of inverse KS formulations. The key observation is that the inverse KS problem is governed by two defining relations: the KS state equations and the density-reproduction condition. Major inversion formulations may then be classified according to whether these
two relations are assigned asymmetric roles, residualized jointly, or imposed
together as feasibility constraints. In this sense,
Wu--Yang appears as a potential-space reduced multiplier formulation \cite{WuYang2003},
ZMP as a quadratic-penalty relaxation \cite{ZhaoMorrisonParr1994,PenzCsirikLaestadius2023},
and PDE-constrained approaches as explicit state-constraint formulations at
the orbital level \cite{JensenWasserman2018,HinzePinnauUlbrichUlbrich2009,Troeltzsch2010}. The same structural perspective also naturally suggests augmented-Lagrangian enrichments, which interpolate between exact multiplier enforcement and pure penalty regularization, as well as enlarged all-at-once formulations in which state consistency and density reproduction are treated within a broader coupled optimization architecture. It also accommodates limiting feasibility-type formulations in which both defining relations are treated primarily as constraints.

Analytical difficulties including noninteracting $v$-representability, nonuniqueness up to constants, nonsmooth variational structure, weak-gap instability, and discretization-induced ill-conditioning remain central in the inversion literature \cite{Capelle2006,Lammert2007,ShiWasserman2021,PenzTellgrenCsirikRuggenthalerLaestadius2023PartI,Herbst2025}. 
The present work organizes these issues at two complementary levels.
Section~II summarizes the associated exact variational and representability questions, while Secs.~III--IV develop an optimization-theoretic formulation map. The map makes explicit which relations are imposed as constraints, dualized, penalized, or residualized, and where metric and gauge choices enter numerical realization, thereby providing a common language for comparing major inversion approaches and guiding algorithmic development.

\section{Variational Foundation of the Inverse Kohn--Sham Problem}

A common heuristic description of inverse Kohn--Sham theory treats the problem as the inversion of a nonlinear forward map
\[
\mathcal M: v \mapsto \rho[v].
\]
This notation is useful at an intuitive level, but it should be understood only schematically, in a regular regime where the ground-state density associated with $v$ is well defined. Its domain and range are not meant to coincide with the full ambient potential and density spaces, and the potential is understood only up to the usual additive-constant ambiguity. More importantly, this picture suppresses the variational structure of exact density functional theory. In particular, it obscures the fact that the effective Kohn--Sham potential arises naturally from the fixed-density noninteracting constrained search, rather than merely as the output of an abstract inverse map \cite{Levy1979,Lieb1983}.

In this section, we place the familiar physical interpretation of inverse Kohn--Sham theory within the function-space constrained-search architecture of exact DFT. The inverse problem then appears as the fixed-density noninteracting constrained search whose value defines $T_\mathrm{s}[\rho]$, making its connections to multiplier theory, regularity, and density--potential duality explicit.

Throughout this work, we use $\langle \cdot,\cdot\rangle$ for single-particle inner products and natural dual pairings, while Dirac bra--ket notation $\langle \cdot|\hat A|\cdot\rangle$ is reserved for many-body operator expectation values. The precise meaning is determined by context. We write $\rho$ for a generic density, $\rho_{\mathrm{tar}}$ for a prescribed target density, and $v$ for a generic multiplier or dual potential variable. Once this dual variable is identified with the effective Kohn--Sham potential, we denote it by $v_\mathrm{s}$.

\subsection{Function spaces and \texorpdfstring{$N$}{N}-representability}

Let $\Omega \subseteq \mathbb R^3$ denote the spatial domain. Depending on the physical setting, one may take $\Omega=\mathbb R^3$ for isolated systems or a periodic torus for extended systems. We write $H^1(\Omega)$ for the usual Sobolev space
\begin{equation}
H^1(\Omega)=\left\{\phi\in L^2(\Omega)\mid \nabla \phi \in [L^2(\Omega)]^3 \right\},
\end{equation}
which provides the natural single-particle energy space for KS orbitals.

For the density variable, we adopt the standard Lieb setting \cite{Lieb1983}. The natural Banach space is
\begin{equation}
X=L^1(\Omega)\cap L^3(\Omega),
\end{equation}
equipped with the norm
\begin{equation}
\|\rho\|_X=\|\rho\|_{L^1}+\|\rho\|_{L^3}.
\end{equation}
Its dual may be identified with
\begin{equation}
X^*=L^\infty(\Omega)+L^{3/2}(\Omega),
\end{equation}
which serves as a natural ambient space for external and effective scalar potentials \cite{Lieb1983}. In the present work, this dual space is used not only for external potentials in the sense of Lieb's formulation, but also as the natural ambient space for the dual variables associated with the noninteracting kinetic functional. In particular, the KS potential $v_\mathrm{s}$ is most naturally viewed as an element of $X^*$ at the level of variational duality, even though additional assumptions are needed to identify such a dual element with a regular local Schr\"odinger potential.

A physically admissible electron density must satisfy more than integrability. We therefore consider densities in the set
\begin{equation}
\mathcal I_N
:=
\left\{
\rho\in X \;\middle|\;
\rho\ge 0,\ 
\int_\Omega \rho = N,\ 
\nabla \sqrt{\rho}\in [L^2(\Omega)]^3
\right\}.
\end{equation}
Here and below, inequalities such as $\rho\ge 0$ are understood almost everywhere. The condition $\nabla\sqrt{\rho}\in L^2$ is the familiar von Weizs\"acker-type regularity requirement. It controls finiteness of the kinetic energy and is closely tied to the existence of finite-energy states reproducing the prescribed density \cite{Lieb1983}.

For sufficiently regular densities in the usual fixed-integer-$N$ domain, Harriman-type constructions establish pure-state, indeed Slater, $N$-representability \cite{Harriman1981,Lieb1983}. Thus the more restrictive issue for inverse KS theory is not merely whether a density can be represented by a Slater determinant, but whether it is the ground-state density of a local noninteracting potential, namely whether it is noninteracting $v$-representable.

At the state level, the corresponding Slater determinants are constructed from orbitals in Sobolev-type energy spaces, while the density variable belongs to the Banach space $X$. This coupling between density-space and orbital-space descriptions is one source of the analytical subtlety of the inverse KS problem \cite{Lieb1983,PenzTellgrenCsirikRuggenthalerLaestadius2023PartI}.

The Lieb $X/X^*$ setting adopted here is not the only possible functional-analytic framework for DFT or inverse KS theory. Alternative Sobolev-type choices, including $H^1/H^{-1}$ frameworks, and Moreau--Yosida regularized settings have also been studied \cite{Sutter2023,Corso2025Rigorous,CorsoLaestadius2025Quantitative,PenzHerbstHelgakerLaestadius2025MY}. Such choices provide additional metric structure and may soften some pathologies of the bare Lieb setting. In the present work, the Lieb setting serves as the standard convex-dual reference framework for the exact variational discussion, whereas particular numerical realizations may introduce additional metrics and regularization.

\subsection{The inverse KS map and its scope}

At the structural level, the inverse KS problem is meaningful only for densities that are noninteracting \(v\)-representable in the relevant class. Most practical inversion algorithms take this representability as an assumption and focus on how to recover the corresponding potential. The variational discussion below makes this assumption explicit, because it is the condition under which the inverse target exists as an exact object. In a regular regime, this may be viewed schematically as a map
\[
\rho \mapsto v_\mathrm{s}[\rho],
\]
defined on an appropriate representable class and only modulo the usual additive-constant ambiguity.

Operationally, the correspondence is defined through the single-particle eigenvalue problem
\begin{equation}
\left(
-\frac{1}{2}\nabla^2+v_\mathrm{s}[\rho](\mathbf r)
\right)\phi_i(\mathbf r)
=
\varepsilon_i \phi_i(\mathbf r),
\qquad i=1,\dots,N,
\label{eq:ks_state}
\end{equation}
together with the density reconstruction condition
\begin{equation}
\rho(\mathbf r)=\sum_{i=1}^N |\phi_i(\mathbf r)|^2.
\label{eq:density_reproduction}
\end{equation}
Thus the inverse problem may be viewed operationally as the search for a local potential whose ground-state orbitals reproduce a prescribed density.

The mathematical status of this correspondence is subtler than the schematic notation suggests. In standard ground-state DFT, the cleanest Hohenberg--Kohn uniqueness statements are obtained in nondegenerate settings, while degeneracy and related nonuniqueness phenomena require more careful treatment of the density--potential relation and its domain of validity \cite{HohenbergKohn1964,Capelle2006,Lammert2007,PenzTellgrenCsirikRuggenthalerLaestadius2023PartI}. Inverse KS theory inherits these subtleties, and practical inversions are further complicated by ill-posedness and instability in finite discretizations \cite{ShiWasserman2021,JensenWasserman2018}.

Accordingly, the discussion below is understood primarily in the regular zero-temperature, fixed-$N$, local-potential regime in which a sufficiently well-behaved inverse KS correspondence may be meaningfully discussed. Our aim is not to settle the full existence theory of the map $\rho\mapsto v_\mathrm{s}[\rho]$, but to explain why, in the regime where such a map is meaningful, the effective KS potential arises naturally from the multiplier structure of the noninteracting constrained search.

\subsection{Fixed-density noninteracting constrained search and the multiplier origin of \texorpdfstring{$v_\mathrm{s}$}{vs}}

At the single-Slater level, let $\Phi$ denote an admissible Slater determinant constructed from orthonormal spin orbitals, with spin variables suppressed in spatial-density formulas. For a prescribed target density $\rho_{\mathrm{tar}}\in \mathcal I_N$, the corresponding determinant-restricted constrained search is
\begin{equation}
T_\mathrm{s}[\rho_{\mathrm{tar}}]
=
\inf_{\Phi \mapsto \rho_{\mathrm{tar}}}
\langle \Phi | \hat T | \Phi \rangle,
\label{eq:Ts_rhotar}
\end{equation}
where the infimum is taken over such determinants with density $\rho_\Phi=\rho_{\mathrm{tar}}$. Thus inverse KS theory is not, at core, an unconstrained search for a potential. It is the study of the constrained minimization problem that defines the noninteracting kinetic energy at a prescribed density \cite{Levy1979,Lieb1983}.

To analyze \eqref{eq:Ts_rhotar}, one introduces the density constraint explicitly and studies the associated Lagrangian. Writing the constraint as $\rho_\Phi=\rho_{\mathrm{tar}}$ and imposing orbital orthonormality gives, formally,
\begin{equation}
\begin{aligned}
\mathcal L(\Phi,v_\mathrm{s},\Lambda)
:={}&
\langle \Phi|\hat T|\Phi\rangle
+
\langle v_\mathrm{s},\rho_\Phi-\rho_{\mathrm{tar}}\rangle \\
&-
\sum_{i,j=1}^N
\Lambda_{ij}\bigl(\langle \phi_i,\phi_j\rangle-\delta_{ij}\bigr),
\end{aligned}
\end{equation}
where $v_\mathrm{s}$ is the multiplier associated with the density constraint and $\Lambda$ collects the orbital orthonormality multipliers.

Stationarity with respect to the orbitals yields the familiar Euler--Lagrange structure
\begin{equation}
\left(
-\frac12\Delta + v_\mathrm{s}(\mathbf r)
\right)\phi_i(\mathbf r)
=
\sum_j \Lambda_{ij}\phi_j(\mathbf r),
\end{equation}
which, after diagonalization of the Hermitian matrix $\Lambda$, becomes
\begin{equation}
\left(
-\frac12\Delta + v_\mathrm{s}(\mathbf r)
\right)\phi_i(\mathbf r)
=
\varepsilon_i \phi_i(\mathbf r).
\end{equation}
Thus the inverse problem may be read in a precise variational sense: one fixes the density and recovers the multiplier field associated with constrained minimization of the noninteracting kinetic energy.

This interpretation is conceptually important. It shows that the effective potential of inverse KS theory is not an ad hoc reconstruction variable introduced from outside the variational structure of DFT. It is the multiplier naturally attached to the fixed-density constrained search \eqref{eq:Ts_rhotar}. In this way, the inverse problem already appears as an intrinsic variational object before one passes to any particular inversion algorithm.

At the same time, this multiplier picture also clarifies why inverse KS theory is analytically delicate. The density constraint is posed in a Banach-space setting appropriate to admissible densities, whereas the state variable $\Phi$ lives in a Sobolev-type energy space. The resulting Banach--Sobolev coupling is precisely what makes the constrained search mathematically nontrivial and what underlies representability, regularity, and stability issues in practice \cite{Lieb1983,PenzTellgrenCsirikRuggenthalerLaestadius2023PartI}.

\subsection{Levy--Lieb theory and the variational position of inverse KS theory}

The multiplier origin of $v_\mathrm{s}$ acquires its full meaning only when placed in the broader setting of the Levy--Lieb constrained-search formulation \cite{Levy1979,Lieb1983}. In that framework, the exact ground-state energy is written as
\begin{equation}
E_0
=
\inf_{\rho\in \mathcal I_N}
\left\{
F_{\mathrm{LL}}[\rho]
+
\langle v_{\mathrm{ext}},\rho\rangle
\right\},
\end{equation}
where
\begin{equation}
F_{\mathrm{LL}}[\rho]
=
\inf_{\Psi \mapsto \rho}
\langle \Psi | \hat T+\hat W | \Psi \rangle.
\end{equation}
Thus the interacting many-body problem already has the form of an outer variational problem over densities together with an inner constrained search over states compatible with those densities.

The fixed-density noninteracting minimization problem in Eq.~\eqref{eq:Ts_rhotar} is not a separate construction specific to inversion. Rather, it is precisely the fixed-density instance of the noninteracting constrained search defining the KS kinetic functional,
\begin{equation}
T_\mathrm{s}[\rho]
=
\inf_{\Phi\mapsto \rho}
\langle \Phi | \hat T | \Phi \rangle.
\end{equation}
For inverse KS theory, one studies this constrained search at the prescribed density $\rho=\rho_{\mathrm{tar}}$ and analyzes the associated multiplier structure.

The KS construction preserves the Levy--Lieb variational logic while reorganizing the universal functional into noninteracting, Hartree, and exchange--correlation parts. Accordingly, one may write the forward KS problem as
\begin{equation}
E_0
=
\inf_{\rho\in \mathcal I_N}
\left\{
\inf_{\Phi\mapsto \rho}
\langle \Phi|\hat T|\Phi\rangle
+
E_\mathrm{H}[\rho]
+
E_\mathrm{xc}[\rho]
+
\langle v_\mathrm{ext},\rho\rangle
\right\}.
\end{equation}
Thus the forward problem already contains an outer--inner variational structure: the outer level selects the energetically optimal density, while the inner level identifies the noninteracting state of minimal kinetic energy compatible with that density.

In a sufficiently regular regime, combining outer stationarity with the inner multiplier structure yields the standard KS relation
\begin{equation}
v_\mathrm{s}[\rho](\mathbf r)
=
v_{\mathrm{ext}}(\mathbf r)
+
v_\mathrm{H}[\rho](\mathbf r)
+
v_{\mathrm{xc}}[\rho](\mathbf r),
\end{equation}
modulo the usual additive constant ambiguity. In this way, the effective potential appearing as a multiplier in the inner constrained search is identified with the usual KS potential of the forward theory.

From this perspective, the inverse KS problem is conceptually very natural. In the forward theory, the density is determined by the outer minimization and the inner constrained search is carried implicitly inside $T_\mathrm{s}[\rho]$. In the inverse problem, by contrast, the density is prescribed in advance, and one studies directly the fixed-density constrained search defining $T_\mathrm{s}[\rho]$ at $\rho=\rho_{\mathrm{tar}}$. The inverse problem is therefore not an artificial optimization overlay imposed on a nonvariational theory. It is the direct study of the multiplier structure of the inner noninteracting constrained search already present in KS theory itself.

For the purposes of the present work, the constrained-search problem \eqref{eq:Ts_rhotar} serves as the common prototype underlying the inversion formulations discussed later. We use $\Phi$ for the noninteracting many-electron state when the constrained-search structure is emphasized, and $\{\phi_i\}_{i=1}^N$ when the orbital-level KS equations are written explicitly.

\subsection{Regularity, representability, and the subdifferential of \texorpdfstring{$T_\mathrm{s}$}{Ts}}

The variational picture developed above is conceptually clean, but it relies on nontrivial assumptions. In particular, it presupposes that the constrained search is attained in a sufficiently regular class, that the target density lies in a regime where a meaningful noninteracting variational description is available, and that the associated variational objects are regular enough for the multiplier interpretation to make classical sense \cite{Lieb1983,PenzTellgrenCsirikRuggenthalerLaestadius2023PartI,Herbst2025}. These issues are closely tied to the long-recognized subtleties of representability, differentiability, and density--potential structure in DFT \cite{Capelle2006,Lammert2007}.

For convex-analytic purposes, it is useful to regard the determinant-restricted orbital search above through its ensemble-extended, lower-semicontinuous convex counterpart; the corresponding density--potential dual construction is developed explicitly in Sec.~II.G. For notational simplicity, we use the same symbol $T_\mathrm{s}$ for this extension in the present subsection. In general, convexity does not imply differentiability \cite{Lieb1983,Rockafellar1970,BauschkeCombettes2011}. Even when $T_\mathrm{s}$ is well defined on a physically relevant domain, it need not be differentiable relative to the fixed-$N$ density domain at every density. This distinction is fundamental for inversion: if $T_\mathrm{s}$ fails to be differentiable at a target density, then the naive identification of the inverse KS potential with a unique classical derivative $\delta T_\mathrm{s}/\delta \rho$ is no longer justified \cite{Lammert2007,PenzTellgrenCsirikRuggenthalerLaestadius2023PartI}. This should be distinguished from formal discussions of functional derivatives of noninteracting kinetic-energy density functionals in smoother or more regular settings \cite{LiuAyers2004}, where the central issue is not the possible failure of differentiability at the target density itself.

A generalized replacement, when available, is provided by the subdifferential. For an extended-valued functional $G$, its effective domain is
\[
\operatorname{dom}G
=
\left\{
\rho \;\middle|\; G[\rho]<+\infty
\right\}.
\]
If $T_\mathrm{s}$ is treated as a proper lower-semicontinuous convex functional on the chosen density space, then for $\rho_0\in\operatorname{dom}T_\mathrm{s}$ one may define
\begin{equation}
\partial T_\mathrm{s}[\rho_0]
=
\left\{
\xi\in X^*
\;\middle|\;
T_\mathrm{s}[\rho]
\ge
T_\mathrm{s}[\rho_0]
+
\langle \xi,\rho-\rho_0\rangle
\quad
\forall \rho\in X
\right\}.
\end{equation}

Because the effective domain of $T_\mathrm{s}$ is contained in the fixed-particle-number affine hyperplane, differentiability should be understood relative to that affine domain. In a regular regime, the derivative is unique as an element of the quotient dual space $X^*/\mathbb R$, rather than as a literal singleton in $X^*$. At points of genuine nondifferentiability, the image of $\partial T_\mathrm{s}[\rho_0]$ in $X^*/\mathbb R$ may contain multiple inequivalent elements \cite{Rockafellar1970,BauschkeCombettes2011}. Finiteness of $T_\mathrm{s}[\rho_0]$ does not by itself imply that $\partial T_\mathrm{s}[\rho_0]$ is nonempty.

In the present setting, the relevance of this construction is direct. The regular multiplier picture suggests that the effective KS potential should be associated, up to sign convention and additive constants, with the variational object dual to the density variable in the constrained search for $T_\mathrm{s}$. When $T_\mathrm{s}$ is differentiable relative to the fixed-$N$ domain, this dual object is represented by a single equivalence class modulo constants. More generally, whenever the subdifferential is nonempty, the relevant dual variational object is an element of $\partial T_\mathrm{s}[\rho_0]$ rather than a classical derivative. Accordingly, the subdifferential framework should be viewed not as separate from the multiplier picture, but as its natural convex-analytic extension beyond the fully regular differentiable regime \cite{PenzTellgrenCsirikRuggenthalerLaestadius2023PartI,PenzCsirikLaestadius2023,Herbst2025}.

This distinction is closely tied to representability. The constrained-search value requires noninteracting $N$-representability: the existence of an admissible Slater determinant in the determinant-restricted formulation, or an admissible ensemble in its convex extension, with density $\rho$. Recovering a local inverse KS potential requires the stronger condition that the density admit a supporting noninteracting potential, equivalently that the relevant subdifferential or multiplier set be nonempty. Thus a density may belong to the domain of the noninteracting constrained search without being noninteracting $v$-representable in the local-potential class. This is distinct from the interacting $v$-representability issue largely bypassed by the Levy--Lieb constrained-search and Lieb convex formulations \cite{Levy1979,Lieb1983,deSilva2012,PenzTellgrenCsirikRuggenthalerLaestadius2023PartI}.

These observations also help explain several familiar numerical pathologies in practical inversion methods, including basis-set artifacts, oscillatory potentials, and instability in weak-gap regimes \cite{ShiWasserman2021,GaidukRyabinkinStaroverov2013,ErhardTrushinGorling2022}. In practice, difficulties often arise when the target density is obtained from incomplete basis representations, strongly correlated states, near-degenerate systems, or limiting regimes in which the noninteracting description becomes fragile. In such cases, inversion procedures that implicitly assume a smooth and single-valued density-to-potential map may become unstable, produce strongly oscillatory effective potentials, or fail to converge robustly.

Differentiability of $T_\mathrm{s}$ should therefore be viewed as a structural assumption rather than a generic property. In favorable regimes, the constrained search gives rise to a well-behaved local multiplier that is unique modulo constants. Outside such regimes, one may instead encounter inequivalent subgradients, loss of smooth duality or dual attainment, or the need for regularized surrogate formulations. These issues provide the analytical background for Sec.~III: they determine when multiplier recovery is available and help explain why penalty or tracking formulations remain meaningful as regularized or approximate problems when exact feasibility or smooth dual structure is unavailable.

\subsection{Gauge structure, local regularity, and asymptotic normalization}

As discussed above, the inverse KS problem determines the effective potential only modulo an additive constant. The natural mathematical object is therefore not a single potential $v_\mathrm{s}$, but an equivalence class
\[
[v_\mathrm{s}]\in X^*/\mathbb R.
\]
This leaves a residual gauge question: once inverse KS theory determines the potential only up to an additive constant, how should the remaining constant be fixed? For finite Coulomb systems, the exact asymptotic structure of the many-body density and of the KS solution selects a distinguished representative from the class $[v_\mathrm{s}]$ and thereby removes the residual gauge freedom \cite{LevyPerdewSahni1984,vanLeeuwenBaerends1994,AlmbladhVonBarth1985}.

Within a sufficiently regular noninteracting $v$-representable class, one expects the inverse correspondence to be injective modulo constants, in the same Hohenberg--Kohn sense that two distinct potentials differing by more than a constant should not generate the same admissible ground-state density \cite{HohenbergKohn1964,Capelle2006,PenzTellgrenCsirikRuggenthalerLaestadius2023PartI}. By contrast, one should not expect global surjectivity on naive ambient spaces such as $X$ and $X^*$. Not every admissible density need arise from a sufficiently regular local noninteracting potential, and not every dual-space potential belongs to the image of a well-behaved inverse KS construction \cite{PenzTellgrenCsirikRuggenthalerLaestadius2023PartI}.

The issue of local regularity is subtler still. Even within representable classes, the dependence of $v_\mathrm{s}[\rho]$ on $\rho$ should be viewed as at most locally regular in favorable regimes. Near degeneracies, gap closings, or boundaries of noninteracting representability, differentiability may fail and the inverse correspondence may cease to be single-valued or stable in the classical sense \cite{ShiWasserman2021,PenzTellgrenCsirikRuggenthalerLaestadius2023PartI,Herbst2025}. It is precisely for this reason that the quotient-valued character of the potential and the need for asymptotic normalization become structurally important rather than merely conventional.

For a fixed external potential, the exchange-correlation potential may be viewed pointwise as the derived quantity
\[
v_{\mathrm{xc}}[\rho](\mathbf r)
=
v_\mathrm{s}[\rho](\mathbf r)-v_{\mathrm{ext}}(\mathbf r)-v_\mathrm{H}[\rho](\mathbf r).
\]
Accordingly, at fixed $v_{\mathrm{ext}}$, whatever regularity or nonuniqueness affects the inverse map $\rho\mapsto v_\mathrm{s}[\rho]$ is inherited, modulo the explicit Hartree contribution, by the exchange-correlation map $\rho\mapsto v_{\mathrm{xc}}[\rho]$.

For finite Coulomb systems, the residual additive ambiguity may be removed by imposing the exact asymptotic structure of the KS solution. Consider a finite $N$-electron Coulomb system with total nuclear charge $Z$, external potential satisfying $v_{\mathrm{ext}}(r)\sim -Z/r$, and exact ground-state density $\rho(r)$. Assume that the system is bound with first ionization energy $I>0$ and that the exact density is noninteracting $v$-representable by a KS potential $v_\mathrm{s}[\rho]$, unique up to an additive constant. In the asymptotic region, away from nodal directions, the exact many-body density has the form
\begin{equation}
\rho(r)\sim C\,r^{2\beta}e^{-2\kappa r},
\qquad
\kappa=\sqrt{2I},
\qquad
\beta=\frac{Z-N+1}{\kappa}-1
\end{equation}
\cite{LevyPerdewSahni1984,AlmbladhVonBarth1985}.

On the other hand, if the KS potential has the asymptotic form $v_\mathrm{s}[\rho](r)\sim -\gamma/r$, then the associated highest occupied orbital obeys
\begin{equation}
\varepsilon_{\mathrm{HOMO}}=-\frac{\kappa^2}{2},
\qquad
\beta=\frac{\gamma}{\kappa}-1.
\end{equation}
Using the ionization-potential theorem $\varepsilon_{\mathrm{HOMO}}=-I$ \cite{PerdewLevy1983,AlmbladhVonBarth1985} and comparing the two expressions for $\beta$, one obtains
\begin{equation}
v_\mathrm{s}[\rho](r)\sim -\frac{Z-N+1}{r},
\qquad r\to\infty.
\end{equation}
Since $v_{\mathrm{ext}}(r)\sim -Z/r$ and $v_\mathrm{H}[\rho](r)\sim N/r$, it follows that
\begin{equation}
v_{\mathrm{xc}}[\rho](r)=-\frac{1}{r}+\mathcal O(r^{-2}),
\qquad r\to\infty.
\end{equation}

This asymptotic analysis has a direct gauge-theoretic interpretation. In a regular noninteracting $v$-representable regime, the inverse KS problem determines only the equivalence class $[v_\mathrm{s}]\in X^*/\mathbb R$. For finite Coulomb systems, however, the exact large-$r$ asymptotic condition selects a distinguished representative from that class. In this sense, asymptotics does not merely provide additional physical information; it resolves the residual gauge indeterminacy of the inverse map and promotes a quotient-valued density-to-potential correspondence to a normalized representative \cite{LevyPerdewSahni1984,vanLeeuwenBaerends1994}.

\subsection{A Lieb-type dual perspective on the noninteracting problem}

The preceding discussion may be reorganized in direct analogy with Lieb's convex formulation of DFT \cite{Lieb1983}. In the interacting case, one passes from the constrained-search functional of Levy and Lieb to the convex-analytic density functional naturally paired with the ground-state energy through Fenchel duality. More precisely, if
\begin{equation}
E[v]
=
\inf_{\rho\in X}
\left\{
F[\rho]+\langle v,\rho\rangle
\right\},
\end{equation}
then the corresponding interacting density functional may be written formally as
\begin{equation}
F[\rho]
=
\sup_{v\in X^*}
\left\{
E[v]-\langle v,\rho\rangle
\right\}.
\end{equation}
Here $F$ denotes the interacting Lieb functional in the convex-analytic sense, namely the lower-semicontinuous convex functional paired with the ground-state energy through Fenchel duality \cite{Lieb1983}. It should be distinguished from the Levy--Lieb constrained-search functional $F_{\mathrm{LL}}$ introduced earlier: while $F_{\mathrm{LL}}$ is defined directly by constrained minimization over interacting wavefunctions, the Lieb functional $F$ is the lower-semicontinuous convex envelope, equivalently the biconjugate, associated with the dual ground-state energy formulation \cite{Lieb1983,Rockafellar1970,LewinLiebSeiringer2022}.

For the convex-dual discussion, it is useful to pass from the determinant-restricted constrained search of Secs.~II.C--II.D to its ensemble extension. Let $\Gamma$ denote an admissible noninteracting ensemble state. The ensemble-extended kinetic functional and noninteracting ground-state energy are
\begin{equation}
T_\mathrm{s}[\rho]
=
\inf_{\Gamma\mapsto\rho}
\operatorname{Tr}(\Gamma\hat T)
\end{equation}
and
\begin{equation}
E_\mathrm{s}[v]
=
\inf_{\Gamma}
\left\{
\operatorname{Tr}(\Gamma\hat T)
+
\langle v,\rho_\Gamma\rangle
\right\},
\label{eq:Es_def}
\end{equation}
where $v$ lies in a suitable potential space dual to the density space. Formally, this yields the density-side variational principle
\begin{equation}
E_\mathrm{s}[v]
=
\inf_{\rho\in X}
\left\{
T_\mathrm{s}[\rho]+\langle v,\rho\rangle
\right\},
\label{eq:Es_dual}
\end{equation}
with $T_\mathrm{s}$ now denoting the ensemble-extended noninteracting density functional. Conversely, in the same convex-analytic spirit,
\begin{equation}
T_\mathrm{s}[\rho]
=
\sup_{v\in X^*}
\left\{
E_\mathrm{s}[v]-\langle v,\rho\rangle
\right\},
\label{eq:Ts_dual}
\end{equation}
up to the usual sign convention for the conjugate pairing \cite{Lieb1983,Rockafellar1970,BauschkeCombettes2011}. With the sign convention in Eqs.~\eqref{eq:Es_dual}--\eqref{eq:Ts_dual}, the central density--potential relation is
\begin{equation}
\rho_{\mathrm{tar}}\in\partial^+E_\mathrm{s}[v_\mathrm{s}]
\quad\Longleftrightarrow\quad
-v_\mathrm{s}\in\partial T_\mathrm{s}[\rho_{\mathrm{tar}}],
\label{eq:subgradient_potential}
\end{equation}
where $\partial^+$ denotes the superdifferential of the concave functional $E_\mathrm{s}$. Because the effective density domain is contained in the affine hyperplane $\int\rho=N$, adding a constant to $v_\mathrm{s}$ leaves the subgradient inequality unchanged on that domain. The usual additive-constant gauge is therefore the dual manifestation of the fixed-particle-number constraint. In this sense, the ensemble extension of the noninteracting kinetic functional admits a Lieb-type dual interpretation closely parallel to that of the interacting functional.

Equivalently, the ensemble constrained search can be expressed in one-particle density-matrix notation,
\begin{equation}
T_\mathrm{s}[\rho]
=
\inf_{\gamma\mapsto \rho}
\operatorname{Tr}\!\left(-\frac12\nabla^2\gamma\right),
\qquad
0\le \gamma\le 1,
\qquad
\operatorname{Tr}\gamma=N,
\end{equation}
with \(\rho_\gamma(\mathbf r)=\sum_\sigma\gamma(\mathbf r\sigma,\mathbf r\sigma)\), with the spin coordinate suppressed elsewhere. This form emphasizes the convex ensemble structure of the exact density-side noninteracting problem. The determinant-restricted constrained search is recovered by imposing idempotency, $\gamma^2=\gamma$, equivalently by restricting to occupied-orbital projectors with integer occupations. When a supporting local potential exists, this specialization yields the usual single-Slater inverse KS problem. In terms of the many-body density matrix, the determinant restriction corresponds to rank-one projectors onto Slater determinants; in terms of the one-particle density matrix, it corresponds to a rank-$N$ idempotent projector.

When an appropriate supporting potential exists at \(\rho_\mathrm{tar}\), the constrained-search problem has a multiplier for the density constraint. The resulting Euler--Lagrange equations are the ensemble KS equations
\begin{equation}
\left(-\frac12\nabla^2+v_\mathrm{s}\right)\phi_i
=
\varepsilon_i\phi_i,
\qquad
\sum_i f_i|\phi_i|^2=\rho_\mathrm{tar},
\end{equation}
with occupation numbers \(0\le f_i\le 1\) and \(\sum_i f_i=N\). At zero temperature, these occupations also satisfy the usual ground-state filling and complementarity conditions; fractional occupations arise only within an appropriate degenerate frontier subspace. The familiar pure-state or single-Slater inverse KS equations are the integer-occupation specialization. Thus the operational inverse KS equations arise from the exact noninteracting density-functional duality after one passes from the ensemble constrained search to an orbital realization.

This dual perspective clarifies the mathematical status of the density--potential mapping. In a sufficiently regular, nondegenerate regime, one may use the schematic notation $\mathcal M:v\mapsto \rho[v]$ and its formal inverse as convenient shorthand. More generally, however, the relevant density--potential relations should be understood through subdifferentials on the density side and superdifferentials, or equivalently subdifferentials of the negated energy functionals, on the potential side \cite{Rockafellar1970,BauschkeCombettes2011,PenzTellgrenCsirikRuggenthalerLaestadius2023PartI}. When the relevant functional is differentiable relative to the fixed-$N$ density domain, the associated differential object reduces to a single equivalence class modulo constants, and the corresponding density--potential relation is locally single-valued in the quotient space. At points of nondifferentiability, by contrast, the relevant differential object remains set-valued even after quotienting out the additive-constant freedom, and the corresponding relation must be understood in a multivalued sense. A distinct failure mode occurs when the density or potential under consideration does not belong to the effective domain of the corresponding functional at all. In that case, one does not obtain a multivalued relation, but rather no density--potential relation is defined there within the given variational framework. Thus the schematic notation $\mathcal M$ and its formal inverse are best regarded as shorthand valid only in sufficiently regular, nondegenerate, and representable regimes.

A further structural issue is that the interacting and noninteracting dual frameworks need not have identical effective domains and ranges. A density that is admissible on the interacting side need not automatically belong to the density class compatible with the noninteracting dual structure. From this viewpoint, the noninteracting $v$-representability problem may be understood as a compatibility problem between the density class accessible in exact DFT and the density class admitting realization within the noninteracting density--potential duality associated with $T_\mathrm{s}$ \cite{Lieb1983,PenzTellgrenCsirikRuggenthalerLaestadius2023PartI}. This should not, however, be read as a claim that the full KS decomposition inherits the same global convex structure as the Lieb functional framework. The relevant convex-analytic structure attaches to the noninteracting component $T_\mathrm{s}$, while the Hartree term is explicit and the exchange-correlation contribution is not in general expected to define a convex functional. The point is therefore not that full KS theory reduces to a globally clean convex program, but rather that its noninteracting component inherits a Lieb-type dual structure that provides the appropriate variational setting for the inverse problem.

\section{Variational Formulations of Inverse Kohn--Sham Theory}

Section~II established the exact variational foundation of inverse KS theory through the noninteracting constrained search and the associated \(T_\mathrm{s}\)--\(E_\mathrm{s}\) density--potential duality. Here we take the resulting orbital relations as the starting point for computational optimization. At the single-Slater level, the operational inverse KS problem is to find orbitals, orbital energies, and a local potential such that
\begin{align}
\left(-\frac12\nabla^2+v\right)\phi_i
&=
\varepsilon_i\phi_i,
\label{eq:feas_state}\\
\sum_{i=1}^N |\phi_i|^2
&=
\rho_\mathrm{tar},
\label{eq:feas_density}\\
\langle\phi_i,\phi_j\rangle
&=
\delta_{ij}.
\label{eq:feas_orth}
\end{align}
The first two equations are the defining state and density-reproduction relations of the inverse KS problem, while orbital orthonormality specifies the admissible orbital manifold. Together they provide a common anchor for the formulations below.

Against this background, three principal realizations are considered in Sec.~III. Multiplier-based reduced formulations originate from a hard density constraint and handle the state dependence through a fixed-potential inner problem; quadratic-penalty formulations relax density reproduction; and explicit state-constraint formulations retain the KS equations as visible constraints while treating density mismatch through a tracking objective. The limiting formulation in which both defining relations are imposed directly as coupled constraints is discussed separately as a feasibility class in Sec.~IV.

The exact and computational levels should nevertheless be distinguished. Noninteracting \(N\)-representability is the feasibility condition for the constrained search defining \(T_\mathrm{s}[\rho]\), whereas recovery of a local inverse KS potential requires the stronger existence of a multiplier or supporting potential, namely noninteracting \(v\)-representability in the relevant potential class. Likewise, convexity of the density-side ensemble functional does not make the explicit orbital or Slater-determinant realization convex: the latter remains a nonconvex problem on an orthonormal orbital manifold. The formulations below therefore describe different optimization-theoretic organizations of the same inverse-KS relations, while their concrete numerical realizations additionally depend on the chosen representation, metric, regularization, and gauge convention.

For notational convenience, when a symmetric positive kernel \(K\) is used, we write
\begin{equation}
\|f\|_K^2
:=
\iint_{\Omega\times\Omega}
f(\mathbf r)\,
K(\mathbf r,\mathbf r')\,
f(\mathbf r')\,
d\mathbf r\,d\mathbf r'.
\end{equation}
When \(K\) is only positive semidefinite, this notation denotes the corresponding seminorm. In orbital-based formulations, we also write
\begin{equation}
\begin{aligned}
D[\{\phi_i\}]
&:=
\sum_{i=1}^N |\phi_i|^2-\rho_{\mathrm{tar}},\\
R_i[\phi_i,v,\varepsilon_i]
&:=
\left(-\frac12\nabla^2+v-\varepsilon_i\right)\phi_i.
\end{aligned}
\end{equation}
Unless otherwise specified, the norms used for state and density residuals are schematic and need not coincide. In continuum formulations, they identify the residual spaces in which the problem is posed; after discretization, their concrete realization depends on the chosen basis, quadrature, and preconditioning. These metric choices are therefore part of the numerical formulation and may also encode regularization.

\subsection{Exact multiplier formulations: Wu--Yang, reduced potential optimization, and duality}

Operationally, the Wu--Yang method may be understood as the reduced-potential formulation obtained from an exact-multiplier treatment of the density-reproduction condition, with the KS state equations handled implicitly through the fixed-potential inner problem. For each fixed trial potential, one solves the inner noninteracting problem subject to orbital orthonormality and obtains a reduced functional of the potential \cite{WuYang2003,KadantsevStott2004,ShiWasserman2021,JensenWasserman2018}. In the language of the Lieb-type dual perspective developed at the end of Sec.~II, Wu--Yang may be viewed as the reduced-potential realization of the noninteracting density--potential dual structure associated with $T_\mathrm{s}$.

Introducing a multiplier $v(\mathbf r)$ for the condition $\rho_\Phi=\rho_{\mathrm{tar}}$ leads formally to the constrained Lagrangian
\begin{equation}
\mathcal L(\Phi,v)
=
\langle \Phi|\hat T|\Phi\rangle
+
\langle v,\rho_\Phi-\rho_{\mathrm{tar}}\rangle,
\end{equation}
supplemented by the orbital orthonormality constraints. This already has the structure of a saddle problem,
\begin{equation}
\inf_{\Phi}\sup_v \mathcal L(\Phi,v).
\end{equation}

The characteristic Wu--Yang reduction is obtained by partitioning the saddle problem into a fixed-potential inner minimization and an outer maximization over the potential. Fix a multiplier field $v\in X^*$ and minimize the Lagrangian over noninteracting states subject to orthonormality. Equivalently, solve the KS state equations for the current trial potential and absorb the state variables into a reduced potential functional. This yields
\begin{equation}
\begin{aligned}
W[v]
:={}&
\inf_{\Phi}
\left\{
\langle \Phi|\hat T|\Phi\rangle
+
\langle v,\rho_\Phi-\rho_{\mathrm{tar}}\rangle
\right\} \\
={}&
E_\mathrm{s}[v]-\langle v,\rho_{\mathrm{tar}}\rangle,
\end{aligned}
\end{equation}
where
\begin{equation}
E_\mathrm{s}[v]
=
\inf_\Phi
\left\{
\langle \Phi|\hat T|\Phi\rangle
+
\langle v,\rho_\Phi\rangle
\right\}
\end{equation}
denotes the noninteracting ground-state energy in the potential $v$. At the exact convex-dual level, the same ground-state energy value is obtained if the infimum is taken over admissible noninteracting ensembles $\Gamma$. In the regular nondegenerate regime underlying the usual single-Slater Wu--Yang realization, a minimizing Slater determinant $\Phi$ suffices; in degenerate regimes, the ensemble formulation and superdifferential description become essential for representing convex combinations of ground-state densities. The Wu--Yang reduced problem is then the maximization
\begin{equation}
\sup_{v\in X^*/\mathbb R} W[v],
\end{equation}
where the quotient indicates the additive-constant gauge freedom.

This reduced functional is exactly what one expects from the noninteracting Lieb-type dual perspective of Sec.~II. There, $E_\mathrm{s}[v]$ was identified as the potential-side object dual to the density functional $T_\mathrm{s}[\rho]$. From this viewpoint, the Wu--Yang functional is simply the potential-space reduced multiplier functional obtained by pairing the noninteracting dual energy with the prescribed target density. In this sense, the Wu--Yang formulation is not merely an ad hoc reduction to potential space, but the most direct reduced-potential expression of the density-centered $T_\mathrm{s}$ framework developed in Sec.~II.

At this point it is useful to distinguish between the reduced-potential viewpoint and the stronger language of exact duality. The reduced functional $W[v]$ is obtained from the hard density-constrained formulation by solving the fixed-potential inner problem and reducing to the potential variable, and this already suffices to explain the operational content of the Wu--Yang algorithm. However, the reduced problem is not automatically identical to the original constrained problem. In general, one only has the weak-duality relation
\begin{equation}
\sup_v \inf_\Phi \mathcal L(\Phi,v)
\;\le\;
\inf_\Phi \sup_v \mathcal L(\Phi,v).
\end{equation}
This is the elementary weak-duality inequality for a saddle formulation: taking the supremum over the multiplier after the infimum over the state variables cannot exceed taking the infimum after the supremum. For the determinant-restricted orbital saddle problem, weak duality is the only automatic statement. After passage to the ensemble-extended lower-semicontinuous convex functional, Fenchel biconjugacy yields equality of the optimal values as in Eq.~\eqref{eq:Ts_dual}. This value equality does not imply that the supremum is attained by a potential in the required local-potential class. Such attainment, equivalently the existence of a supporting potential satisfying Eq.~\eqref{eq:subgradient_potential}, is the additional noninteracting $v$-representability or dual-attainment requirement relevant to inverse KS reconstruction. Thus the reduced maximization over $v$ should not be identified automatically with the determinant-restricted density-constrained problem without additional assumptions ensuring an appropriate saddle structure, dual attainment, constraint qualification, or strong-duality property \cite{Lieb1983,Rockafellar1970,BauschkeCombettes2011}.

Assuming sufficient regularity of the noninteracting ground-state energy with respect to the potential, one has
\begin{equation}
\frac{\delta E_\mathrm{s}[v]}{\delta v(\mathbf r)}=\rho_v(\mathbf r),
\end{equation}
where $\rho_v$ is the noninteracting ground-state density associated with $v$. Hence
\begin{equation}
\frac{\delta W[v]}{\delta v(\mathbf r)}
=
\rho_v(\mathbf r)-\rho_{\mathrm{tar}}(\mathbf r).
\end{equation}
A stationary point of $W[v]$ therefore corresponds formally to exact satisfaction of the density-reproduction condition.

Under nondegeneracy and suitable spectral-gap assumptions, the second variation of $W[v]$ is governed by the static noninteracting density-response operator,
\begin{equation}
\frac{\delta^2 W}{\delta v(\mathbf r)\,\delta v(\mathbf r')}
=
\chi_\mathrm{s}(\mathbf r,\mathbf r'),
\end{equation}
where, in a standard orbital representation,
\begin{equation}
\chi_\mathrm{s}(\mathbf r,\mathbf r')
=
2\operatorname{Re}
\sum_{i\in \mathrm{occ}}
\sum_{a\in \mathrm{unocc}}
\frac{
\phi_i^*(\mathbf r)\phi_a(\mathbf r)\phi_a^*(\mathbf r')\phi_i(\mathbf r')
}{
\varepsilon_i-\varepsilon_a
}.
\end{equation}
The constant-potential direction is an unavoidable null mode of $\chi_\mathrm{s}$, since an additive shift changes the orbital energies but not the orbitals or density. Newton-type updates must therefore be formulated after gauge fixing or directly on the quotient potential space. This makes the local optimization structure transparent: the Wu--Yang functional naturally supports Newton or quasi-Newton optimization in potential space, with gradient given by density mismatch and Hessian governed by the KS response operator. It also explains why the method becomes fragile in small-gap or nearly degenerate regimes, where the same response operator becomes poorly conditioned \cite{ShiWasserman2021,JensenWasserman2018,ErhardTrushinGorling2022,Gould2023}.

\textit{Structural conclusion.} The Wu--Yang formulation is the potential-space reduced formulation obtained from the density-constrained noninteracting problem: the orbitals are solved for each trial potential, and the resulting functional \(W[v]=E_\mathrm{s}[v]-\langle v,\rho_{\mathrm{tar}}\rangle\) is maximized over the potential modulo the additive-constant gauge.

\subsection{Quadratic-penalty relaxations: the ZMP paradigm}

A natural alternative to exact multiplier enforcement is to relax the density-reproduction condition through a quadratic penalty term. Operationally, the KS state equations remain in place as the forward state relation associated with the current effective potential, but exact density matching is no longer imposed as a hard constraint. Instead, density mismatch is assigned a quadratic cost in the objective. From the present viewpoint, this is the essential variational structure of the Zhao--Morrison--Parr method \cite{ZhaoMorrisonParr1994}. In other words, ZMP is best understood as a quadratic-penalty relaxation of the density-constrained inverse KS problem \cite{ZhaoMorrisonParr1994,PenzCsirikLaestadius2023,ShiWasserman2021}.

At the schematic level, one replaces the exact constrained problem by a penalized functional of the form
\begin{equation}
\mathcal J_\lambda[\Phi]
=
\langle \Phi|\hat T|\Phi\rangle
+
\frac{\lambda}{2}\|\rho_\Phi-\rho_{\mathrm{tar}}\|^2,
\end{equation}
or, more generally, by a kernel-weighted quadratic penalty
\begin{equation}
\mathcal J_\lambda[\Phi]
=
\langle \Phi|\hat T|\Phi\rangle
+
\frac{\lambda}{2}\|\rho_\Phi-\rho_{\mathrm{tar}}\|_K^2.
\end{equation}
Here $K$ is a symmetric positive kernel and $\lambda>0$ is the penalty parameter. In the original ZMP construction, the Coulomb kernel is the canonical choice \cite{ZhaoMorrisonParr1994}, but the structural point is more general: the hard density constraint has been replaced by a quadratic penalty that drives the solution toward the target density as $\lambda$ increases.

Thus ZMP does not solve the hard-constrained inverse problem exactly at finite $\lambda$; rather, it solves a penalized surrogate problem. Accordingly, the potential obtained at finite \(\lambda\) should be viewed as
a penalty-surrogate potential, not by itself as the exact inverse KS potential. For feasible states satisfying $\rho_\Phi=\rho_{\mathrm{tar}}$, the penalty term vanishes, so the penalized objective agrees with the exact constrained objective on the constraint manifold. Away from that manifold, however, the minimizer may trade density mismatch against kinetic optimality. ZMP should therefore be understood as a family of approximate inverse problems that approach the exact density-constrained problem only in the large-penalty limit.

The corresponding effective potential is obtained by taking the first variation of the penalty term with respect to the density. One is then led to the density-dependent correction
\begin{equation}
v_\lambda(\mathbf r)
=
\lambda\,(K*(\rho_\Phi-\rho_{\mathrm{tar}}))(\mathbf r).
\end{equation}
Thus the density mismatch is fed back directly into the one-body potential through the kernel $K$. This is the characteristic operational content of ZMP: rather than introducing an independent multiplier variable and solving a reduced optimization problem over potentials, one iteratively corrects the effective potential by penalizing the current density residual \cite{ZhaoMorrisonParr1994,ShiChavezWasserman2022}.

The practical attraction of the penalty viewpoint is robustness. Even when the target density lies near the boundary of regular noninteracting $v$-representability, the penalized problem remains meaningful as an approximate inversion task. One no longer asks for exact feasibility at every stage, but for the best compromise between noninteracting kinetic optimality and density matching at finite $\lambda$. This frequently makes the optimization more forgiving than an exact reduced-multiplier or reduced-dual formulation \cite{ShiWasserman2021,ErhardTrushinGorling2022,Gould2023}.

At the same time, the weakness of quadratic-penalty methods is classical. If one wants to recover the exact density constraint, the penalty parameter must be taken large. As $\lambda$ increases, however, the geometry of the orbital optimization becomes increasingly anisotropic. Let $\mathcal T[\Phi]=\langle\Phi|\hat T|\Phi\rangle$ and $C[\Phi]=\rho_\Phi-\rho_{\mathrm{tar}}$. Near feasibility, a schematic Gauss--Newton-type approximation on the tangent space of the orthonormal orbital manifold is
\begin{equation}
\nabla_\Phi^2 \mathcal J_\lambda
\approx
\nabla_\Phi^2 \mathcal T
+
\lambda C'[\Phi]^*K C'[\Phi].
\end{equation}
The exact Hessian also contains a residual-weighted term coupling $C''[\Phi]$ to $K C[\Phi]$, which vanishes at exact feasibility and is neglected in this local approximation. The displayed term makes clear that density-changing directions acquire stiffness proportional to $\lambda$, while directions weakly seen by $K^{1/2}C'[\Phi]$ do not \cite{Rockafellar1970,BauschkeCombettes2011}.

In practical implementations, one often increases the penalty parameter progressively. The resulting family of penalized problems therefore admits a continuation, or homotopy-like, interpretation: one follows a path from relatively soft density matching toward increasingly accurate enforcement of the target density as $\lambda$ grows. More abstractly, the ZMP construction lies close in spirit to Moreau--Yosida-type regularization \cite{PenzCsirikLaestadius2023}. The hard density constraint is replaced by a smoother family of penalized problems in which exact feasibility is traded for improved local regularity of the objective landscape. The precise extent to which a given ZMP formulation coincides with a classical Moreau envelope depends on the function-space setting and metric structure, but this convex-analytic interpretation is best understood as a posterior explanation of the quadratic-penalty structure rather than as the primary definition of the method.

Like Wu--Yang, this formulation remains closely tied to the density-centered constrained-search picture, but replaces exact multiplier enforcement by quadratic penalization.

\textit{Structural conclusion.} The ZMP formulation is the quadratic-penalty relaxation of the same density-constrained inverse Kohn--Sham problem.

\subsection{Explicit state-constraint formulations: the PDE-constrained paradigm}

A third route is to enlarge the constrained system rather than to reduce the state dependence through a fixed-potential inner problem or to relax the density condition through a quadratic penalty alone. In the tracking variant considered here, the KS state equations \eqref{eq:ks_state} are retained explicitly as constraints, while the density-reproduction condition \eqref{eq:density_reproduction} enters through a tracking-type objective. Formulations that impose both relations as hard constraints belong instead to the coupled-feasibility class discussed in Sec.~IV.C. Here ``tracking'' is used in the standard PDE-constrained optimization sense: the objective measures the mismatch between a state-dependent quantity and a prescribed target. From the present viewpoint, this is the essential structure of the PDE-constrained paradigm: the inverse KS problem is formulated as an explicit state-constraint optimization problem rather than as a reduced multiplier problem or a pure penalty relaxation \cite{JensenWasserman2018,HinzePinnauUlbrichUlbrich2009,Troeltzsch2010,ShiWasserman2021}.

At the schematic level, one considers an optimization problem of the form
\begin{equation}
\min_{\{\phi_i\},\,v,\,\{\varepsilon_i\}}\; \mathcal J[\{\phi_i\}]
\end{equation}
subject to the KS state equations \eqref{eq:ks_state} and the usual orbital orthonormality conditions. Here $\{\phi_i\}_{i=1}^N$ are the state variables, $v$ is the control variable, $\{\varepsilon_i\}$ are auxiliary state-equation unknowns, and $\mathcal J$ is typically a density-mismatch functional such as
\begin{equation}
\mathcal J[\{\phi_i\}]
=
\frac12 \|\rho-\rho_{\mathrm{tar}}\|_K^2,
\qquad
\rho=\sum_{i=1}^N |\phi_i|^2.
\end{equation}
In this example, the tracked quantity is the density. The quadratic density-tracking objective is the least-squares realization of the density-reproduction condition once the KS equations are retained as hard state constraints, and it selects a metric on density residuals. In an attainable regime, zero-residual minimizers recover exact density matching; otherwise, the formulation identifies the closest representable density in the chosen metric. The metric is therefore part of the numerical realization rather than an intrinsic feature of exact DFT.

Equivalently, if $C[\{\phi_i\}]=\rho-\rho_{\mathrm{tar}}$, then exact inversion corresponds to $C=0$, whereas the tracking formulation minimizes $\frac12\|C\|_K^2$. The Euler--Lagrange equations of the tracking problem do not impose $C=0$ as a hard constraint; instead, the density residual enters the adjoint equation through the variational derivative of the objective. Thus density matching is treated as an objective-level zero-residual condition, while the KS equations remain explicit state constraints.

Unlike Wu--Yang, this approach does not reduce the problem to a potential-only outer optimization. Unlike ZMP, it does not derive the state equations from a density-penalized orbital functional. Instead, it treats the orbitals, orbital energies, and potential within an explicit state--control formulation.

To derive first-order optimality conditions, one introduces adjoint variables $\{p_i\}_{i=1}^N$ and forms a control Lagrangian
\begin{equation}
\begin{aligned}
\mathcal L(\{\phi_i\},v,\{\varepsilon_i\},\{p_i\},\Lambda)
:={}&
\mathcal J[\{\phi_i\}] \\
&+
\sum_{i=1}^N
\left\langle
p_i,\,
R_i[\phi_i,v,\varepsilon_i]
\right\rangle \\
&+
\sum_{i,j=1}^N
\Lambda_{ij}\bigl(\langle \phi_i,\phi_j\rangle-\delta_{ij}\bigr),
\end{aligned}
\end{equation}
where $\Lambda$ enforces orbital orthonormality. The resulting stationarity conditions couple the state equations, adjoint equations, normalization conditions, and control gradient relation into a single optimality system \cite{HinzePinnauUlbrichUlbrich2009,Troeltzsch2010}. For notational economy, we reuse the symbol $\mathcal L$ for the relevant Lagrangian in each formulation.

The adjoint equations arise from variation with respect to the orbitals and have the characteristic form of projected Sternheimer or linear-response equations. For the quadratic tracking objective above, the density residual enters as the adjoint source. If $q=K(\rho-\rho_\mathrm{tar})$, then schematically,
\begin{equation}
\left(
-\frac12\nabla^2+v(\mathbf r)-\varepsilon_k
\right)
p_k(\mathbf r)
=
-2q(\mathbf r)\phi_k(\mathbf r)
-
\sum_{j=1}^N \Lambda_{kj}\phi_j(\mathbf r),
\end{equation}
to be understood in the appropriate weak or projected sense. The essential computational point is that, once the adjoint system is solved, the gradient of the objective with respect to the control variable may be obtained without explicitly differentiating the orbitals with respect to the potential \cite{JensenWasserman2018,HinzePinnauUlbrichUlbrich2009,Troeltzsch2010}.

The resulting control gradient is obtained by variation with respect to the potential. Formally, it has the schematic structure
\begin{equation}
\frac{\delta \mathcal L}{\delta v(\mathbf r)}
\sim
2\operatorname{Re}\sum_{k=1}^N \phi_k^*(\mathbf r)p_k(\mathbf r),
\end{equation}
up to the precise pairing convention and weak-form normalization being used. Thus the inversion may be attacked by gradient-based optimization in the control variable, with derivative information supplied by state--adjoint pairs. Under linearization of the density map, the corresponding reduced stationarity condition has the normal-equation form $\chi_\mathrm{s}^*K(\rho[v]-\rho_\mathrm{tar})=0$. This is why the metric used in the density-tracking term and the response properties of the KS map directly control the conditioning of the PDE-constrained formulation.

At the same time, the weakness of the explicit state-constraint paradigm is equally clear. The projected adjoint equations involve shifted KS operators whose inverses are controlled by spectral gaps. In well-gapped systems, the reduced operator behaves stably and the adjoint equations provide a reliable route to accurate gradients. In weak-gap, nearly degenerate, or metallic regimes, however, the same projected operators become poorly conditioned or nearly singular. The resulting loss of robustness is therefore not accidental; it reflects the spectral structure of the underlying linearized inverse problem itself \cite{JensenWasserman2018,ShiWasserman2021,ErhardTrushinGorling2022,Gould2023}.

Thus the PDE-constrained formulation is most naturally understood at the
broader structural level, where the KS state equations are retained explicitly
rather than absorbed into a \(T_\mathrm{s}\)-centered reduced formulation.

\textit{Structural conclusion.} PDE-constrained inversion constitutes the explicit state-constraint realization of the inverse Kohn--Sham problem at the orbital level.

Response-based and stabilized inversion schemes can often be interpreted as
algorithmic realizations, regularized variants, or iterative solvers associated
with the same underlying structural classes identified here, rather than as
fundamentally separate variational or optimization-theoretic classes.

\section{Optimization-theoretic classification of inverse Kohn--Sham formulations}

Having identified three principal realizations in Sec.~III, we now abstract the common optimization pattern. The primary taxonomy is governed by the roles assigned to the two defining inverse-KS relations---the KS state equations and the density-reproduction condition. It distinguishes asymmetric formulations, in which the two relations are reduced, dualized, penalized, constrained, or tracked differently; joint-residual formulations, in which both relations enter a scalar objective through separate residual blocks; and coupled feasibility formulations, in which both are imposed as constraints. The same architecture may be stated at the level of exact continuum variational relations, formal constrained-optimization problems, or numerical realizations specified by a representation, residual norm, metric, preconditioner, and gauge convention; equivalence across these levels requires the corresponding regularity, attainment, and constraint-qualification assumptions \cite{JensenWasserman2018,HinzePinnauUlbrichUlbrich2009,Troeltzsch2010}.

A complementary distinction concerns the optimization variables that remain explicit. Reduced formulations solve the state variables through an inner problem and optimize over a smaller variable block, whereas state--control and all-at-once formulations retain the orbitals, potential, and, where appropriate, orbital energies as coupled unknowns. This variable-elimination structure is independent of the relation-role taxonomy below. Orbital orthonormality remains present throughout but is treated separately because it defines the admissible orbital manifold rather than either target inverse-KS relation; it may be enforced by Lagrange multipliers or absorbed by working on a Stiefel or Grassmann manifold.

\subsection{Asymmetric formulations: different roles for the two defining relations}

The first class assigns different optimization roles to the two defining inverse-KS relations. One relation may be retained explicitly, while the other is handled through an inner solve, a multiplier or dual formulation, a penalty, or a tracking objective. This class contains the main paradigms discussed in Sec.~III.

On the density-centered side, the fixed-density constrained search retains density reproduction as a hard constraint and obtains the KS equations from stationarity. Wu--Yang solves the fixed-potential state problem internally and reduces the formulation to potential space \cite{WuYang2003}; ZMP instead replaces exact density reproduction by a quadratic penalty \cite{ZhaoMorrisonParr1994,PenzCsirikLaestadius2023}. Conversely, PDE-constrained inversion retains the KS equations as explicit state constraints and measures density mismatch through a tracking objective over the coupled state--control variables \cite{JensenWasserman2018,HinzePinnauUlbrichUlbrich2009,Troeltzsch2010}.

Within this class, augmented-Lagrangian ideas enrich a hard density constraint by combining multiplier information with moderate regularization \cite{Rockafellar1970,BauschkeCombettes2011}.

Consider the density-centered constrained-search viewpoint of Sec.~II,
\begin{equation}
\min_{\Phi}\; \langle \Phi|\hat T|\Phi\rangle
\qquad
\text{subject to }
\rho_\Phi=\rho_{\mathrm{tar}}.
\end{equation}
An augmented-Lagrangian enrichment of this formulation takes the form
\begin{equation}
\mathcal L[\Phi,\mu]
=
\langle \Phi|\hat T|\Phi\rangle
+
\langle \mu,\rho_\Phi-\rho_{\mathrm{tar}}\rangle
+
\frac{\lambda}{2}\|\rho_\Phi-\rho_{\mathrm{tar}}\|^2,
\end{equation}
where \(\mu\in X^*\) is the multiplier associated with the density-reproduction constraint and \(\lambda>0\) is a finite penalty parameter. If the quadratic term is removed, one recovers the exact-multiplier viewpoint underlying Wu--Yang; if the multiplier is frozen or omitted, one recovers a ZMP-type penalty structure. In this sense, the augmented-Lagrangian construction is a natural extension within the same asymmetric class.

Relative to a pure exact-multiplier formulation, the quadratic term provides additional local stabilization and may improve robustness of the constrained search. Relative to a pure penalty formulation, the multiplier absorbs part of the burden of feasibility, so that exact density reproduction need not be approached solely through a numerically dangerous large-penalty limit. At the level of optimization structure, this viewpoint yields a natural
primal--dual coupled formulation. An Uzawa iteration is a standard strategy
for equality-constrained optimization: one alternates between approximately
minimizing the augmented Lagrangian with respect to the primal variables and
updating the multiplier in the direction of the constraint residual. In the
present inverse-KS setting, the primal variable is the noninteracting state or
orbital set, and the multiplier update is driven by the density mismatch
after it has been mapped into the chosen multiplier space. Algorithmically,
this suggests iterations such as
\begin{equation}
\Phi^{(k+1)}
\approx
\arg\min_{\Phi}\,
\mathcal L\big[\Phi,\mu^{(k)}\big],
\end{equation}
followed by
\begin{equation}
\mu^{(k+1)}
=
\mu^{(k)}
+
\lambda\,\mathcal P\!\left(\rho_{\Phi^{(k+1)}}-\rho_{\mathrm{tar}}\right).
\end{equation}
Here \(\mathcal P\) is a chosen residual-to-multiplier map from the density
residual space to the potential-like dual space. In a Hilbert or
finite-dimensional discretized setting, \(\mathcal P\) may be the Riesz map,
a quadrature-weighted mass-matrix inverse, a Coulomb kernel, a Sobolev
preconditioner, or another regularizing operator. In the bare Lieb
\(X/X^*\) setting there is no canonical identification of density residuals
with potentials, so this update should be understood as a metric-dependent
numerical realization rather than a literal Banach-space identity. Residual-driven
potential updates of this general type are closely related in spirit to the
van Leeuwen--Baerends inversion scheme, where the density error is used to
iteratively correct the potential \cite{vanLeeuwenBaerends1994}.

\subsection{Joint residual formulations}

A second class places both defining inverse-KS relations into a single scalar objective through separate residual blocks. This leads to enlarged all-at-once formulations in which state consistency and density reproduction are treated simultaneously without designating either relation as the unique retained outer constraint.

Writing the occupied orbitals as $\{\phi_i\}_{i=1}^N$, one natural example is
\begin{equation}
\min_{\{\phi_i\},\,v,\,\{\varepsilon_i\}}
\;
w_1\sum_{i=1}^N \|R_i[\phi_i,v,\varepsilon_i]\|^2
+
w_2\|D[\{\phi_i\}]\|^2,
\end{equation}
together with the usual orbital orthonormality constraints and, where desired, additional regularity penalties on the potential or gauge-fixing terms. Here the first term measures violation of the KS state equations, while the second measures violation of density reproduction. In such a formulation, neither relation is treated solely as a hidden solver nor solely as a hard constraint. Instead, both enter symmetrically as coupled optimization targets.

This enlarged formulation unifies structures that otherwise appear separately
in reduced potential formulations, quadratic-penalty relaxations, and
explicit state-constraint methods. At the same time, it introduces
nontrivial design choices: one must decide how the two residual blocks
should be balanced, whether through fixed weights $w_1,w_2$, continuation in
one or both weights, or adaptive merit strategies that respond to the
evolving scale of the state and density residuals.

At the algorithmic level, this opens a substantially richer design space than the classical nested inversion picture. One could imagine alternating minimization in $(\{\phi_i\},v,\{\varepsilon_i\})$, block-coordinate descent, sequential quadratic programming, Newton--Krylov methods on the coupled residual system, or hybrid schemes in which adjoint-based derivative information is combined with multiplier updates and continuation in regularization parameters \cite{HinzePinnauUlbrichUlbrich2009,Troeltzsch2010}. Likewise, modern parameterized representations of orbitals and potentials, such as finite-dimensional basis expansions, symmetry-adapted ans\"atze, and neural-network or physics-informed neural-network (PINN)-style parameterizations, may permit genuinely joint optimization strategies that are difficult to formulate naturally within rigidly asymmetric inversion formulations and schemes \cite{RaissiPerdikarisKarniadakis2019PINN}.

Residualizing both defining inverse-KS relations in a joint scalar objective defines
a distinct optimization class. Its stationary points depend on the residual
norms, weights, and parameterization and need not correspond to feasible
inverse-KS solutions. This class therefore exposes an algorithmic design space
whose relation to the exact continuum formulation must be assessed for the
chosen residual geometry and representation.

\textit{Structural conclusion.} Enlarged all-at-once residual formulations
constitute the joint-residual class in the optimization-theoretic taxonomy of
the inverse Kohn--Sham problem.

\subsection{Coupled feasibility formulations}

A third limiting class is obtained when both defining inverse-KS relations are treated as constraints rather than as objectives. In this case, the inverse problem is viewed primarily as a coupled nonlinear system to be solved rather than as a conventional optimization problem with a distinguished cost functional. One then seeks orbitals, a potential, and orbital energies satisfying both the KS state equations \eqref{eq:ks_state} and the density-reproduction condition \eqref{eq:density_reproduction} simultaneously, together with the usual orbital orthonormality conditions and, where needed, a gauge-fixing condition for the potential.

This viewpoint may be framed formally as the feasibility problem
\begin{equation}
\text{find }(\{\phi_i\},v,\{\varepsilon_i\})
\quad\text{subject to}\quad
\eqref{eq:ks_state}
\ \text{and}\
\eqref{eq:density_reproduction},
\end{equation}
together with the usual orbital orthonormality conditions. The goal is not to minimize a preferred objective while enforcing one dominant constraint, but to find a jointly consistent solution of the coupled state and density relations.

Even in the absence of a primary outer objective, it is still natural to write down a constrained Lagrangian. Using the same orbital-level notation as above, let $\{p_i\}_{i=1}^N$ denote multipliers associated with the KS state equations, let $\mu(\mathbf r)$ denote the multiplier associated with density reproduction, and let $\Lambda=(\Lambda_{ij})$ enforce orbital orthonormality, while the orbital energies $\{\varepsilon_i\}_{i=1}^N$ are treated as auxiliary unknowns of the explicit state-equation block. One is then led schematically to
\begin{equation}
\begin{aligned}
\mathcal L(\{\phi_i\},v,\{\varepsilon_i\},\{p_i\},\mu,\Lambda)
:={}&
\langle \mu, D[\{\phi_i\}] \rangle \\
&+
\sum_{i=1}^N \langle p_i, R_i[\phi_i,v,\varepsilon_i]\rangle \\
&+
\sum_{i,j=1}^N
\Lambda_{ij}\bigl(\langle\phi_i,\phi_j\rangle-\delta_{ij}\bigr).
\end{aligned}
\end{equation}
In this way, one obtains a formal multiplier or primal--dual representation in which both defining inverse relations appear explicitly as constraints. Because the objective is identically zero, the multiplier system does not by itself select among feasible solutions; its role here is to expose the coupled constraint structure and to motivate primal--dual or equation-solving realizations.

This feasibility viewpoint completes the optimization-theoretic
classification by treating both defining relations as constraints rather than
as objectives or penalties.

\textit{Structural conclusion.} Coupled feasibility formulations constitute
the feasibility class in the optimization-theoretic taxonomy of the inverse
Kohn--Sham problem, with orbital orthonormality understood as the additional
structural constraint defining the orbital manifold.

\section{Conclusion}

We have developed a unified mathematical framework for the inverse
Kohn--Sham problem by proceeding in two steps. First, we identified the
density-constrained noninteracting constrained search associated with exact
density functional theory as the natural variational anchor of the inverse KS problem. Second, building on this anchor, we developed an
optimization-theoretic classification of inverse KS formulations according
to whether the KS state equations and density-reproduction condition are
assigned asymmetric roles, residualized jointly, or imposed as coupled
feasibility constraints.

Within this framework, Wu--Yang appears as a potential-space reduced multiplier formulation, ZMP as a quadratic-penalty relaxation of the same density-constrained problem, and PDE-constrained approaches as explicit state-constraint formulations at the orbital level. This common viewpoint identifies where additive-constant ambiguity, asymptotic normalization, nondifferentiability, loss of dual attainment, metric choice, and weak-gap instability enter different inversion architectures: they reflect different manifestations of the same underlying variational and structural content rather than unrelated algorithmic pathologies \cite{ShiWasserman2021,PenzCsirikLaestadius2023,Herbst2025}.
The point of this classification is therefore structural: it organizes inverse
KS theory at the level of optimization architecture rather than at the level of
particular established algorithms. In particular, it separates the exact feasibility target from the different ways in which the KS equation and density matching may be reduced, constrained, dualized, penalized, tracked, or residualized, while retaining orbital orthonormality as an additional manifold constraint. Systematic convergence analysis and numerical comparison of the resulting
formulation classes provide natural directions for future work.

The framework also naturally accommodates broader formulation classes. In particular, augmented-Lagrangian formulations provide an intermediate route between exact multiplier enforcement and penalty regularization, while enlarged all-at-once formulations offer a setting in which state consistency and density reproduction may be treated on more nearly equal footing. These extensions broaden the optimization-theoretic picture developed here, although their detailed analytical properties and numerical conditioning still merit systematic study.

More broadly, the inverse KS problem may be viewed not only as a practical tool for extracting reference exchange-correlation potentials, but also as a structured model problem at the intersection of density functional theory, convex analysis, nonsmooth optimization, and PDE-constrained control \cite{Lieb1983,Rockafellar1970,BauschkeCombettes2011,HinzePinnauUlbrichUlbrich2009,Troeltzsch2010}. We hope that the present formulation helps sharpen these connections and provides a clearer language for comparing existing inversion formulations and methods and for developing new ones.

\begin{acknowledgments}
The author thanks Wenzhi Gao for helpful discussions.
\end{acknowledgments}

\bibliography{ref}

\begin{thebibliography}{42}%
\makeatletter
\providecommand \@ifxundefined [1]{%
 \@ifx{#1\undefined}
}%
\providecommand \@ifnum [1]{%
 \ifnum #1\expandafter \@firstoftwo
 \else \expandafter \@secondoftwo
 \fi
}%
\providecommand \@ifx [1]{%
 \ifx #1\expandafter \@firstoftwo
 \else \expandafter \@secondoftwo
 \fi
}%
\providecommand \natexlab [1]{#1}%
\providecommand \enquote  [1]{``#1''}%
\providecommand \bibnamefont  [1]{#1}%
\providecommand \bibfnamefont [1]{#1}%
\providecommand \citenamefont [1]{#1}%
\providecommand \href@noop [0]{\@secondoftwo}%
\providecommand \href [0]{\begingroup \@sanitize@url \@href}%
\providecommand \@href[1]{\@@startlink{#1}\@@href}%
\providecommand \@@href[1]{\endgroup#1\@@endlink}%
\providecommand \@sanitize@url [0]{\catcode `\\12\catcode `\$12\catcode `\&12\catcode `\#12\catcode `\^12\catcode `\_12\catcode `\%12\relax}%
\providecommand \@@startlink[1]{}%
\providecommand \@@endlink[0]{}%
\providecommand \url  [0]{\begingroup\@sanitize@url \@url }%
\providecommand \@url [1]{\endgroup\@href {#1}{\urlprefix }}%
\providecommand \urlprefix  [0]{URL }%
\providecommand \Eprint [0]{\href }%
\providecommand \doibase [0]{http://dx.doi.org/}%
\providecommand \selectlanguage [0]{\@gobble}%
\providecommand \bibinfo  [0]{\@secondoftwo}%
\providecommand \bibfield  [0]{\@secondoftwo}%
\providecommand \translation [1]{[#1]}%
\providecommand \BibitemOpen [0]{}%
\providecommand \bibitemStop [0]{}%
\providecommand \bibitemNoStop [0]{.\EOS\space}%
\providecommand \EOS [0]{\spacefactor3000\relax}%
\providecommand \BibitemShut  [1]{\csname bibitem#1\endcsname}%
\let\auto@bib@innerbib\@empty
\bibitem [{\citenamefont {Hohenberg}\ and\ \citenamefont {Kohn}(1964)}]{HohenbergKohn1964}%
  \BibitemOpen
  \bibfield  {author} {\bibinfo {author} {\bibfnamefont {P.}~\bibnamefont {Hohenberg}}\ and\ \bibinfo {author} {\bibfnamefont {W.}~\bibnamefont {Kohn}},\ }\href {\doibase 10.1103/PhysRev.136.B864} {\bibfield  {journal} {\bibinfo  {journal} {Physical Review}\ }\textbf {\bibinfo {volume} {136}},\ \bibinfo {pages} {B864} (\bibinfo {year} {1964})}\BibitemShut {NoStop}%
\bibitem [{\citenamefont {Kohn}\ and\ \citenamefont {Sham}(1965)}]{KohnSham1965}%
  \BibitemOpen
  \bibfield  {author} {\bibinfo {author} {\bibfnamefont {W.}~\bibnamefont {Kohn}}\ and\ \bibinfo {author} {\bibfnamefont {L.~J.}\ \bibnamefont {Sham}},\ }\href {\doibase 10.1103/PhysRev.140.A1133} {\bibfield  {journal} {\bibinfo  {journal} {Physical Review}\ }\textbf {\bibinfo {volume} {140}},\ \bibinfo {pages} {A1133} (\bibinfo {year} {1965})}\BibitemShut {NoStop}%
\bibitem [{\citenamefont {Shi}\ and\ \citenamefont {Wasserman}(2021)}]{ShiWasserman2021}%
  \BibitemOpen
  \bibfield  {author} {\bibinfo {author} {\bibfnamefont {Y.}~\bibnamefont {Shi}}\ and\ \bibinfo {author} {\bibfnamefont {A.}~\bibnamefont {Wasserman}},\ }\href {\doibase 10.1021/acs.jpclett.1c00752} {\bibfield  {journal} {\bibinfo  {journal} {The Journal of Physical Chemistry Letters}\ }\textbf {\bibinfo {volume} {12}},\ \bibinfo {pages} {5308} (\bibinfo {year} {2021})}\BibitemShut {NoStop}%
\bibitem [{\citenamefont {Jensen}\ and\ \citenamefont {Wasserman}(2018)}]{JensenWasserman2018}%
  \BibitemOpen
  \bibfield  {author} {\bibinfo {author} {\bibfnamefont {D.~S.}\ \bibnamefont {Jensen}}\ and\ \bibinfo {author} {\bibfnamefont {A.}~\bibnamefont {Wasserman}},\ }\href {\doibase 10.1002/qua.25425} {\bibfield  {journal} {\bibinfo  {journal} {International Journal of Quantum Chemistry}\ }\textbf {\bibinfo {volume} {118}},\ \bibinfo {pages} {e25425} (\bibinfo {year} {2018})}\BibitemShut {NoStop}%
\bibitem [{\citenamefont {Kanungo}, \citenamefont {Zimmerman},\ and\ \citenamefont {Gavini}(2019)}]{KanungoZimmermanGavini2019}%
  \BibitemOpen
  \bibfield  {author} {\bibinfo {author} {\bibfnamefont {B.}~\bibnamefont {Kanungo}}, \bibinfo {author} {\bibfnamefont {P.~M.}\ \bibnamefont {Zimmerman}}, \ and\ \bibinfo {author} {\bibfnamefont {V.}~\bibnamefont {Gavini}},\ }\href {\doibase 10.1038/s41467-019-12467-0} {\bibfield  {journal} {\bibinfo  {journal} {Nature Communications}\ }\textbf {\bibinfo {volume} {10}},\ \bibinfo {pages} {4497} (\bibinfo {year} {2019})}\BibitemShut {NoStop}%
\bibitem [{\citenamefont {Shi}, \citenamefont {Ch{\'a}vez},\ and\ \citenamefont {Wasserman}(2022)}]{ShiChavezWasserman2022}%
  \BibitemOpen
  \bibfield  {author} {\bibinfo {author} {\bibfnamefont {Y.}~\bibnamefont {Shi}}, \bibinfo {author} {\bibfnamefont {V.~H.}\ \bibnamefont {Ch{\'a}vez}}, \ and\ \bibinfo {author} {\bibfnamefont {A.}~\bibnamefont {Wasserman}},\ }\href {\doibase 10.1002/wcms.1617} {\bibfield  {journal} {\bibinfo  {journal} {WIREs Computational Molecular Science}\ }\textbf {\bibinfo {volume} {12}},\ \bibinfo {pages} {e1617} (\bibinfo {year} {2022})}\BibitemShut {NoStop}%
\bibitem [{\citenamefont {Kanungo}\ \emph {et~al.}(2023)\citenamefont {Kanungo}, \citenamefont {Hatch}, \citenamefont {Zimmerman},\ and\ \citenamefont {Gavini}}]{KanungoZimmermanGavini2023}%
  \BibitemOpen
  \bibfield  {author} {\bibinfo {author} {\bibfnamefont {B.}~\bibnamefont {Kanungo}}, \bibinfo {author} {\bibfnamefont {J.}~\bibnamefont {Hatch}}, \bibinfo {author} {\bibfnamefont {P.~M.}\ \bibnamefont {Zimmerman}}, \ and\ \bibinfo {author} {\bibfnamefont {V.}~\bibnamefont {Gavini}},\ }\href {\doibase 10.1021/acs.jpclett.3c01713} {\bibfield  {journal} {\bibinfo  {journal} {The Journal of Physical Chemistry Letters}\ }\textbf {\bibinfo {volume} {14}},\ \bibinfo {pages} {10039} (\bibinfo {year} {2023})}\BibitemShut {NoStop}%
\bibitem [{\citenamefont {Zhao}, \citenamefont {Morrison},\ and\ \citenamefont {Parr}(1994)}]{ZhaoMorrisonParr1994}%
  \BibitemOpen
  \bibfield  {author} {\bibinfo {author} {\bibfnamefont {Q.}~\bibnamefont {Zhao}}, \bibinfo {author} {\bibfnamefont {R.~C.}\ \bibnamefont {Morrison}}, \ and\ \bibinfo {author} {\bibfnamefont {R.~G.}\ \bibnamefont {Parr}},\ }\href {\doibase 10.1103/PhysRevA.50.2138} {\bibfield  {journal} {\bibinfo  {journal} {Physical Review A}\ }\textbf {\bibinfo {volume} {50}},\ \bibinfo {pages} {2138} (\bibinfo {year} {1994})}\BibitemShut {NoStop}%
\bibitem [{\citenamefont {Wu}\ and\ \citenamefont {Yang}(2003)}]{WuYang2003}%
  \BibitemOpen
  \bibfield  {author} {\bibinfo {author} {\bibfnamefont {Q.}~\bibnamefont {Wu}}\ and\ \bibinfo {author} {\bibfnamefont {W.}~\bibnamefont {Yang}},\ }\href {\doibase 10.1063/1.1535422} {\bibfield  {journal} {\bibinfo  {journal} {The Journal of Chemical Physics}\ }\textbf {\bibinfo {volume} {118}},\ \bibinfo {pages} {2498} (\bibinfo {year} {2003})}\BibitemShut {NoStop}%
\bibitem [{\citenamefont {Kadantsev}\ and\ \citenamefont {Stott}(2004)}]{KadantsevStott2004}%
  \BibitemOpen
  \bibfield  {author} {\bibinfo {author} {\bibfnamefont {E.~S.}\ \bibnamefont {Kadantsev}}\ and\ \bibinfo {author} {\bibfnamefont {M.~J.}\ \bibnamefont {Stott}},\ }\href {\doibase 10.1103/PhysRevA.69.012502} {\bibfield  {journal} {\bibinfo  {journal} {Physical Review A}\ }\textbf {\bibinfo {volume} {69}},\ \bibinfo {pages} {012502} (\bibinfo {year} {2004})}\BibitemShut {NoStop}%
\bibitem [{\citenamefont {Gaiduk}, \citenamefont {Ryabinkin},\ and\ \citenamefont {Staroverov}(2013)}]{GaidukRyabinkinStaroverov2013}%
  \BibitemOpen
  \bibfield  {author} {\bibinfo {author} {\bibfnamefont {A.~P.}\ \bibnamefont {Gaiduk}}, \bibinfo {author} {\bibfnamefont {I.~G.}\ \bibnamefont {Ryabinkin}}, \ and\ \bibinfo {author} {\bibfnamefont {V.~N.}\ \bibnamefont {Staroverov}},\ }\href {\doibase 10.1021/ct4004146} {\bibfield  {journal} {\bibinfo  {journal} {Journal of Chemical Theory and Computation}\ }\textbf {\bibinfo {volume} {9}},\ \bibinfo {pages} {3959} (\bibinfo {year} {2013})}\BibitemShut {NoStop}%
\bibitem [{\citenamefont {Zhang}\ and\ \citenamefont {Carter}(2018)}]{ZhangCarter2018}%
  \BibitemOpen
  \bibfield  {author} {\bibinfo {author} {\bibfnamefont {X.}~\bibnamefont {Zhang}}\ and\ \bibinfo {author} {\bibfnamefont {E.~A.}\ \bibnamefont {Carter}},\ }\href {\doibase 10.1063/1.5005839} {\bibfield  {journal} {\bibinfo  {journal} {The Journal of Chemical Physics}\ }\textbf {\bibinfo {volume} {148}},\ \bibinfo {pages} {034105} (\bibinfo {year} {2018})}\BibitemShut {NoStop}%
\bibitem [{\citenamefont {Erhard}, \citenamefont {Trushin},\ and\ \citenamefont {G{\"o}rling}(2022)}]{ErhardTrushinGorling2022}%
  \BibitemOpen
  \bibfield  {author} {\bibinfo {author} {\bibfnamefont {J.}~\bibnamefont {Erhard}}, \bibinfo {author} {\bibfnamefont {E.}~\bibnamefont {Trushin}}, \ and\ \bibinfo {author} {\bibfnamefont {A.}~\bibnamefont {G{\"o}rling}},\ }\href {\doibase 10.1063/5.0087356} {\bibfield  {journal} {\bibinfo  {journal} {The Journal of Chemical Physics}\ }\textbf {\bibinfo {volume} {156}},\ \bibinfo {pages} {204124} (\bibinfo {year} {2022})}\BibitemShut {NoStop}%
\bibitem [{\citenamefont {Gould}(2023)}]{Gould2023}%
  \BibitemOpen
  \bibfield  {author} {\bibinfo {author} {\bibfnamefont {T.}~\bibnamefont {Gould}},\ }\href {\doibase 10.1063/5.0134330} {\bibfield  {journal} {\bibinfo  {journal} {The Journal of Chemical Physics}\ }\textbf {\bibinfo {volume} {158}},\ \bibinfo {pages} {064102} (\bibinfo {year} {2023})}\BibitemShut {NoStop}%
\bibitem [{\citenamefont {Ryabinkin}\ and\ \citenamefont {Staroverov}(2012)}]{RyabinkinStaroverov2012}%
  \BibitemOpen
  \bibfield  {author} {\bibinfo {author} {\bibfnamefont {I.~G.}\ \bibnamefont {Ryabinkin}}\ and\ \bibinfo {author} {\bibfnamefont {V.~N.}\ \bibnamefont {Staroverov}},\ }\href {\doibase 10.1063/1.4763481} {\bibfield  {journal} {\bibinfo  {journal} {The Journal of Chemical Physics}\ }\textbf {\bibinfo {volume} {137}},\ \bibinfo {pages} {164113} (\bibinfo {year} {2012})}\BibitemShut {NoStop}%
\bibitem [{\citenamefont {Ryabinkin}, \citenamefont {Kohut},\ and\ \citenamefont {Staroverov}(2015)}]{RyabinkinKohutStaroverov2015}%
  \BibitemOpen
  \bibfield  {author} {\bibinfo {author} {\bibfnamefont {I.~G.}\ \bibnamefont {Ryabinkin}}, \bibinfo {author} {\bibfnamefont {S.~V.}\ \bibnamefont {Kohut}}, \ and\ \bibinfo {author} {\bibfnamefont {V.~N.}\ \bibnamefont {Staroverov}},\ }\href {\doibase 10.1103/PhysRevLett.115.083001} {\bibfield  {journal} {\bibinfo  {journal} {Physical Review Letters}\ }\textbf {\bibinfo {volume} {115}},\ \bibinfo {pages} {083001} (\bibinfo {year} {2015})}\BibitemShut {NoStop}%
\bibitem [{\citenamefont {Cuevas-Saavedra}, \citenamefont {Ayers},\ and\ \citenamefont {Staroverov}(2015)}]{CuevasSaavedraAyersStaroverov2015}%
  \BibitemOpen
  \bibfield  {author} {\bibinfo {author} {\bibfnamefont {R.}~\bibnamefont {Cuevas-Saavedra}}, \bibinfo {author} {\bibfnamefont {P.~W.}\ \bibnamefont {Ayers}}, \ and\ \bibinfo {author} {\bibfnamefont {V.~N.}\ \bibnamefont {Staroverov}},\ }\href {\doibase 10.1063/1.4937943} {\bibfield  {journal} {\bibinfo  {journal} {The Journal of Chemical Physics}\ }\textbf {\bibinfo {volume} {143}},\ \bibinfo {pages} {244116} (\bibinfo {year} {2015})}\BibitemShut {NoStop}%
\bibitem [{\citenamefont {Levy}(1979)}]{Levy1979}%
  \BibitemOpen
  \bibfield  {author} {\bibinfo {author} {\bibfnamefont {M.}~\bibnamefont {Levy}},\ }\href {\doibase 10.1073/pnas.76.12.6062} {\bibfield  {journal} {\bibinfo  {journal} {Proceedings of the National Academy of Sciences of the United States of America}\ }\textbf {\bibinfo {volume} {76}},\ \bibinfo {pages} {6062} (\bibinfo {year} {1979})}\BibitemShut {NoStop}%
\bibitem [{\citenamefont {Lieb}(1983)}]{Lieb1983}%
  \BibitemOpen
  \bibfield  {author} {\bibinfo {author} {\bibfnamefont {E.~H.}\ \bibnamefont {Lieb}},\ }\href {\doibase 10.1002/qua.560240302} {\bibfield  {journal} {\bibinfo  {journal} {International Journal of Quantum Chemistry}\ }\textbf {\bibinfo {volume} {24}},\ \bibinfo {pages} {243} (\bibinfo {year} {1983})}\BibitemShut {NoStop}%
\bibitem [{\citenamefont {Sheng}(2026)}]{sheng2026exact}%
  \BibitemOpen
  \bibfield  {author} {\bibinfo {author} {\bibfnamefont {N.}~\bibnamefont {Sheng}},\ }\href {\doibase 10.48550/arXiv.2603.23399} {} (\bibinfo {year} {2026}),\ \Eprint {http://arxiv.org/abs/2603.23399} {arXiv:2603.23399 [physics.chem-ph]} \BibitemShut {NoStop}%
\bibitem [{\citenamefont {Penz}, \citenamefont {Csirik},\ and\ \citenamefont {Laestadius}(2023)}]{PenzCsirikLaestadius2023}%
  \BibitemOpen
  \bibfield  {author} {\bibinfo {author} {\bibfnamefont {M.}~\bibnamefont {Penz}}, \bibinfo {author} {\bibfnamefont {M.~A.}\ \bibnamefont {Csirik}}, \ and\ \bibinfo {author} {\bibfnamefont {A.}~\bibnamefont {Laestadius}},\ }\href {\doibase 10.1088/2516-1075/acc626} {\bibfield  {journal} {\bibinfo  {journal} {Electronic Structure}\ }\textbf {\bibinfo {volume} {5}},\ \bibinfo {pages} {014009} (\bibinfo {year} {2023})}\BibitemShut {NoStop}%
\bibitem [{\citenamefont {Hinze}\ \emph {et~al.}(2009)\citenamefont {Hinze}, \citenamefont {Pinnau}, \citenamefont {Ulbrich},\ and\ \citenamefont {Ulbrich}}]{HinzePinnauUlbrichUlbrich2009}%
  \BibitemOpen
  \bibfield  {author} {\bibinfo {author} {\bibfnamefont {M.}~\bibnamefont {Hinze}}, \bibinfo {author} {\bibfnamefont {R.}~\bibnamefont {Pinnau}}, \bibinfo {author} {\bibfnamefont {M.}~\bibnamefont {Ulbrich}}, \ and\ \bibinfo {author} {\bibfnamefont {S.}~\bibnamefont {Ulbrich}},\ }\href {\doibase 10.1007/978-1-4020-8839-1} {\emph {\bibinfo {title} {Optimization with PDE Constraints}}}\ (\bibinfo  {publisher} {Springer},\ \bibinfo {address} {New York},\ \bibinfo {year} {2009})\BibitemShut {NoStop}%
\bibitem [{\citenamefont {Tr{\"o}ltzsch}(2010)}]{Troeltzsch2010}%
  \BibitemOpen
  \bibfield  {author} {\bibinfo {author} {\bibfnamefont {F.}~\bibnamefont {Tr{\"o}ltzsch}},\ }\href {\doibase 10.1090/gsm/112} {\emph {\bibinfo {title} {Optimal Control of Partial Differential Equations: Theory, Methods and Applications}}}\ (\bibinfo  {publisher} {American Mathematical Society},\ \bibinfo {address} {Providence, RI},\ \bibinfo {year} {2010})\BibitemShut {NoStop}%
\bibitem [{\citenamefont {Capelle}(2006)}]{Capelle2006}%
  \BibitemOpen
  \bibfield  {author} {\bibinfo {author} {\bibfnamefont {K.}~\bibnamefont {Capelle}},\ }\href {\doibase 10.1590/S0103-97332006000700035} {\bibfield  {journal} {\bibinfo  {journal} {Brazilian Journal of Physics}\ }\textbf {\bibinfo {volume} {36}},\ \bibinfo {pages} {1318} (\bibinfo {year} {2006})}\BibitemShut {NoStop}%
\bibitem [{\citenamefont {Lammert}(2007)}]{Lammert2007}%
  \BibitemOpen
  \bibfield  {author} {\bibinfo {author} {\bibfnamefont {P.~E.}\ \bibnamefont {Lammert}},\ }\href {\doibase 10.1002/qua.21342} {\bibfield  {journal} {\bibinfo  {journal} {International Journal of Quantum Chemistry}\ }\textbf {\bibinfo {volume} {107}},\ \bibinfo {pages} {1943} (\bibinfo {year} {2007})}\BibitemShut {NoStop}%
\bibitem [{\citenamefont {Penz}\ \emph {et~al.}(2023)\citenamefont {Penz}, \citenamefont {Tellgren}, \citenamefont {Csirik}, \citenamefont {Ruggenthaler},\ and\ \citenamefont {Laestadius}}]{PenzTellgrenCsirikRuggenthalerLaestadius2023PartI}%
  \BibitemOpen
  \bibfield  {author} {\bibinfo {author} {\bibfnamefont {M.}~\bibnamefont {Penz}}, \bibinfo {author} {\bibfnamefont {E.~I.}\ \bibnamefont {Tellgren}}, \bibinfo {author} {\bibfnamefont {M.~A.}\ \bibnamefont {Csirik}}, \bibinfo {author} {\bibfnamefont {M.}~\bibnamefont {Ruggenthaler}}, \ and\ \bibinfo {author} {\bibfnamefont {A.}~\bibnamefont {Laestadius}},\ }\href {\doibase 10.1021/acsphyschemau.2c00069} {\bibfield  {journal} {\bibinfo  {journal} {ACS Physical Chemistry Au}\ }\textbf {\bibinfo {volume} {3}},\ \bibinfo {pages} {334} (\bibinfo {year} {2023})}\BibitemShut {NoStop}%
\bibitem [{\citenamefont {Herbst}, \citenamefont {Bakkestuen},\ and\ \citenamefont {Laestadius}(2025)}]{Herbst2025}%
  \BibitemOpen
  \bibfield  {author} {\bibinfo {author} {\bibfnamefont {M.~F.}\ \bibnamefont {Herbst}}, \bibinfo {author} {\bibfnamefont {V.~H.}\ \bibnamefont {Bakkestuen}}, \ and\ \bibinfo {author} {\bibfnamefont {A.}~\bibnamefont {Laestadius}},\ }\href {\doibase 10.1103/PhysRevB.111.205143} {\bibfield  {journal} {\bibinfo  {journal} {Physical Review B}\ }\textbf {\bibinfo {volume} {111}},\ \bibinfo {pages} {205143} (\bibinfo {year} {2025})}\BibitemShut {NoStop}%
\bibitem [{\citenamefont {Harriman}(1981)}]{Harriman1981}%
  \BibitemOpen
  \bibfield  {author} {\bibinfo {author} {\bibfnamefont {J.~E.}\ \bibnamefont {Harriman}},\ }\href {\doibase 10.1103/PhysRevA.24.680} {\bibfield  {journal} {\bibinfo  {journal} {Physical Review A}\ }\textbf {\bibinfo {volume} {24}},\ \bibinfo {pages} {680} (\bibinfo {year} {1981})}\BibitemShut {NoStop}%
\bibitem [{\citenamefont {Sutter}\ \emph {et~al.}(2024)\citenamefont {Sutter}, \citenamefont {Penz}, \citenamefont {Ruggenthaler}, \citenamefont {van Leeuwen},\ and\ \citenamefont {Giesbertz}}]{Sutter2023}%
  \BibitemOpen
  \bibfield  {author} {\bibinfo {author} {\bibfnamefont {S.~M.}\ \bibnamefont {Sutter}}, \bibinfo {author} {\bibfnamefont {M.}~\bibnamefont {Penz}}, \bibinfo {author} {\bibfnamefont {M.}~\bibnamefont {Ruggenthaler}}, \bibinfo {author} {\bibfnamefont {R.}~\bibnamefont {van Leeuwen}}, \ and\ \bibinfo {author} {\bibfnamefont {K.~J.~H.}\ \bibnamefont {Giesbertz}},\ }\href {\doibase 10.1088/1751-8121/ad8a2a} {\bibfield  {journal} {\bibinfo  {journal} {Journal of Physics A: Mathematical and Theoretical}\ }\textbf {\bibinfo {volume} {57}},\ \bibinfo {pages} {475202} (\bibinfo {year} {2024})},\ \Eprint {http://arxiv.org/abs/2312.07225} {arXiv:2312.07225 [math-ph]} \BibitemShut {NoStop}%
\bibitem [{\citenamefont {Corso}(2025)}]{Corso2025Rigorous}%
  \BibitemOpen
  \bibfield  {author} {\bibinfo {author} {\bibfnamefont {T.~C.}\ \bibnamefont {Corso}},\ }\href {\doibase 10.48550/arXiv.2504.05501} {} (\bibinfo {year} {2025}),\ \Eprint {http://arxiv.org/abs/2504.05501} {arXiv:2504.05501 [math-ph]} \BibitemShut {NoStop}%
\bibitem [{\citenamefont {Corso}\ and\ \citenamefont {Laestadius}(2025)}]{CorsoLaestadius2025Quantitative}%
  \BibitemOpen
  \bibfield  {author} {\bibinfo {author} {\bibfnamefont {T.~C.}\ \bibnamefont {Corso}}\ and\ \bibinfo {author} {\bibfnamefont {A.}~\bibnamefont {Laestadius}},\ }\href {\doibase 10.48550/arXiv.2512.04726} {} (\bibinfo {year} {2025}),\ \Eprint {http://arxiv.org/abs/2512.04726} {arXiv:2512.04726 [math-ph]} \BibitemShut {NoStop}%
\bibitem [{\citenamefont {Penz}\ \emph {et~al.}(2026)\citenamefont {Penz}, \citenamefont {Herbst}, \citenamefont {Helgaker},\ and\ \citenamefont {Laestadius}}]{PenzHerbstHelgakerLaestadius2025MY}%
  \BibitemOpen
  \bibfield  {author} {\bibinfo {author} {\bibfnamefont {M.}~\bibnamefont {Penz}}, \bibinfo {author} {\bibfnamefont {M.~F.}\ \bibnamefont {Herbst}}, \bibinfo {author} {\bibfnamefont {T.}~\bibnamefont {Helgaker}}, \ and\ \bibinfo {author} {\bibfnamefont {A.}~\bibnamefont {Laestadius}},\ }\href {\doibase 10.1088/2516-1075/ae55eb} {\bibfield  {journal} {\bibinfo  {journal} {Electronic Structure}\ }\textbf {\bibinfo {volume} {8}},\ \bibinfo {pages} {022001} (\bibinfo {year} {2026})},\ \Eprint {http://arxiv.org/abs/2511.06957} {arXiv:2511.06957 [cond-mat.mtrl-sci]} \BibitemShut {NoStop}%
\bibitem [{\citenamefont {Rockafellar}(1970)}]{Rockafellar1970}%
  \BibitemOpen
  \bibfield  {author} {\bibinfo {author} {\bibfnamefont {R.~T.}\ \bibnamefont {Rockafellar}},\ }\href@noop {} {\emph {\bibinfo {title} {Convex Analysis}}}\ (\bibinfo  {publisher} {Princeton University Press},\ \bibinfo {address} {Princeton, NJ},\ \bibinfo {year} {1970})\BibitemShut {NoStop}%
\bibitem [{\citenamefont {Bauschke}\ and\ \citenamefont {Combettes}(2011)}]{BauschkeCombettes2011}%
  \BibitemOpen
  \bibfield  {author} {\bibinfo {author} {\bibfnamefont {H.~H.}\ \bibnamefont {Bauschke}}\ and\ \bibinfo {author} {\bibfnamefont {P.~L.}\ \bibnamefont {Combettes}},\ }\href {\doibase 10.1007/978-1-4419-9467-7} {\emph {\bibinfo {title} {Convex Analysis and Monotone Operator Theory in Hilbert Spaces}}}\ (\bibinfo  {publisher} {Springer},\ \bibinfo {address} {New York},\ \bibinfo {year} {2011})\BibitemShut {NoStop}%
\bibitem [{\citenamefont {Liu}\ and\ \citenamefont {Ayers}(2004)}]{LiuAyers2004}%
  \BibitemOpen
  \bibfield  {author} {\bibinfo {author} {\bibfnamefont {S.}~\bibnamefont {Liu}}\ and\ \bibinfo {author} {\bibfnamefont {P.~W.}\ \bibnamefont {Ayers}},\ }\href {\doibase 10.1103/PhysRevA.70.022501} {\bibfield  {journal} {\bibinfo  {journal} {Physical Review A}\ }\textbf {\bibinfo {volume} {70}},\ \bibinfo {pages} {022501} (\bibinfo {year} {2004})}\BibitemShut {NoStop}%
\bibitem [{\citenamefont {de~Silva}\ and\ \citenamefont {Wesolowski}(2012)}]{deSilva2012}%
  \BibitemOpen
  \bibfield  {author} {\bibinfo {author} {\bibfnamefont {P.}~\bibnamefont {de~Silva}}\ and\ \bibinfo {author} {\bibfnamefont {T.~A.}\ \bibnamefont {Wesolowski}},\ }\href {\doibase 10.1103/PhysRevA.85.032518} {\bibfield  {journal} {\bibinfo  {journal} {Physical Review A}\ }\textbf {\bibinfo {volume} {85}},\ \bibinfo {pages} {032518} (\bibinfo {year} {2012})}\BibitemShut {NoStop}%
\bibitem [{\citenamefont {Levy}, \citenamefont {Perdew},\ and\ \citenamefont {Sahni}(1984)}]{LevyPerdewSahni1984}%
  \BibitemOpen
  \bibfield  {author} {\bibinfo {author} {\bibfnamefont {M.}~\bibnamefont {Levy}}, \bibinfo {author} {\bibfnamefont {J.~P.}\ \bibnamefont {Perdew}}, \ and\ \bibinfo {author} {\bibfnamefont {V.}~\bibnamefont {Sahni}},\ }\href {\doibase 10.1103/PhysRevA.30.2745} {\bibfield  {journal} {\bibinfo  {journal} {Physical Review A}\ }\textbf {\bibinfo {volume} {30}},\ \bibinfo {pages} {2745} (\bibinfo {year} {1984})}\BibitemShut {NoStop}%
\bibitem [{\citenamefont {van Leeuwen}\ and\ \citenamefont {Baerends}(1994)}]{vanLeeuwenBaerends1994}%
  \BibitemOpen
  \bibfield  {author} {\bibinfo {author} {\bibfnamefont {R.}~\bibnamefont {van Leeuwen}}\ and\ \bibinfo {author} {\bibfnamefont {E.~J.}\ \bibnamefont {Baerends}},\ }\href {\doibase 10.1103/PhysRevA.49.2421} {\bibfield  {journal} {\bibinfo  {journal} {Physical Review A}\ }\textbf {\bibinfo {volume} {49}},\ \bibinfo {pages} {2421} (\bibinfo {year} {1994})}\BibitemShut {NoStop}%
\bibitem [{\citenamefont {Almbladh}\ and\ \citenamefont {von Barth}(1985)}]{AlmbladhVonBarth1985}%
  \BibitemOpen
  \bibfield  {author} {\bibinfo {author} {\bibfnamefont {C.-O.}\ \bibnamefont {Almbladh}}\ and\ \bibinfo {author} {\bibfnamefont {U.}~\bibnamefont {von Barth}},\ }\href {\doibase 10.1103/PhysRevB.31.3231} {\bibfield  {journal} {\bibinfo  {journal} {Physical Review B}\ }\textbf {\bibinfo {volume} {31}},\ \bibinfo {pages} {3231} (\bibinfo {year} {1985})}\BibitemShut {NoStop}%
\bibitem [{\citenamefont {Perdew}\ and\ \citenamefont {Levy}(1983)}]{PerdewLevy1983}%
  \BibitemOpen
  \bibfield  {author} {\bibinfo {author} {\bibfnamefont {J.~P.}\ \bibnamefont {Perdew}}\ and\ \bibinfo {author} {\bibfnamefont {M.}~\bibnamefont {Levy}},\ }\href {\doibase 10.1103/PhysRevLett.51.1884} {\bibfield  {journal} {\bibinfo  {journal} {Physical Review Letters}\ }\textbf {\bibinfo {volume} {51}},\ \bibinfo {pages} {1884} (\bibinfo {year} {1983})}\BibitemShut {NoStop}%
\bibitem [{\citenamefont {Lewin}, \citenamefont {Lieb},\ and\ \citenamefont {Seiringer}(2022)}]{LewinLiebSeiringer2022}%
  \BibitemOpen
  \bibfield  {author} {\bibinfo {author} {\bibfnamefont {M.}~\bibnamefont {Lewin}}, \bibinfo {author} {\bibfnamefont {E.~H.}\ \bibnamefont {Lieb}}, \ and\ \bibinfo {author} {\bibfnamefont {R.}~\bibnamefont {Seiringer}},\ }\href {\doibase 10.48550/arXiv.1912.10424} {} (\bibinfo {year} {2022}),\ \Eprint {http://arxiv.org/abs/1912.10424} {arXiv:1912.10424 [math-ph]} \BibitemShut {NoStop}%
\bibitem [{\citenamefont {Raissi}, \citenamefont {Perdikaris},\ and\ \citenamefont {Karniadakis}(2019)}]{RaissiPerdikarisKarniadakis2019PINN}%
  \BibitemOpen
  \bibfield  {author} {\bibinfo {author} {\bibfnamefont {M.}~\bibnamefont {Raissi}}, \bibinfo {author} {\bibfnamefont {P.}~\bibnamefont {Perdikaris}}, \ and\ \bibinfo {author} {\bibfnamefont {G.~E.}\ \bibnamefont {Karniadakis}},\ }\href {\doibase 10.1016/j.jcp.2018.10.045} {\bibfield  {journal} {\bibinfo  {journal} {Journal of Computational Physics}\ }\textbf {\bibinfo {volume} {378}},\ \bibinfo {pages} {686} (\bibinfo {year} {2019})}\BibitemShut {NoStop}%
\end{thebibliography}%

\end{document}